\documentclass[12pt,a4paper,final]{iopart}

\newcommand{\pT}{p_{T}}

\newcommand{\barQ}{{\bar{Q}}}
\newcommand{\barq}{{\bar{q}}}
\newcommand{\barc}{{\bar{c}}}

\newcommand{\cs}{{\hat{s}}}
\newcommand{\ct}{{\hat{t}}}
\newcommand{\cu}{{\hat{u}}}
\newcommand{\alphas}{{\alpha_{s}}}
\newcommand{\shat}{\hat{\rm s}}
\newcommand{\that}{\hat{\rm t}}
\newcommand{\uhat}{\hat{\rm u}}
\newcommand{\zhat}{\hat{\rm z}}
\newcommand{\CA}{{\cal A}}
\newcommand{\Qbar}{{\overline Q}}

\newcommand{\QQbaroctetsingS}{{Q\Qbar[ ^1S_0^{(8)}]}}
\newcommand{\QQbaroctettripP}{{Q\Qbar[ ^3P_J^{(8)}]}}

\def\QQbaroctettripS{Q\Qbar[ ^3S_1^{(8)}]}
\def\QQbaroctetPzero{Q\Qbar[ ^3P_0^{(8)}]}
\def\QQbaroctetPone{Q\Qbar[ ^3P_1^{(8)}]}
\def\QQbaroctetPtwo{Q\Qbar[ ^3P_2^{(8)}]}

\usepackage{verbatim}
\usepackage{graphicx}
\usepackage[usenames,dvipsnames,svgnames,table]{xcolor}
\usepackage[breaklinks=true,colorlinks=true,linkcolor=blue,urlcolor=blue,citecolor=blue]{hyperref}
\usepackage{rotating}
\usepackage{graphicx}
\usepackage{dcolumn}
\usepackage{bm}
\usepackage{epsfig}
\usepackage{hyperref}
\usepackage{ulem}
\usepackage{appendix}
\usepackage{iopams}
\usepackage{mwe}
\usepackage{subfig}
\expandafter\let\csname equation*\endcsname\relax
\expandafter\let\csname endequation*\endcsname\relax
\usepackage{amsmath}

\begin{document}

\title[]{Charmonia production in p+p collisions under NRQCD formalism}

\author{Vineet Kumar}
\address{Nuclear Physics Division, Bhabha Atomic Research Center, Mumbai, India}
\ead{vineet.kumar@cern.ch}
\author{Prashant Shukla$^{1,2}$}
\address{$^{1}$Nuclear Physics Division, Bhabha Atomic Research Center, Mumbai, India}
\address{$^{2}$Homi Bhabha National Institute, Anushakti Nagar, Mumbai, India}
\ead{pshukla@barc.gov.in}


\begin{abstract}
  This work presents the differential charmonia production cross sections in high 
  energy p+p collisions calculated  using NRQCD formalism. The NRQCD formalism, 
  factorizes the quarkonia production cross sections in terms of short distance QCD 
  cross sections and long distance matrix elements (LDMEs). The short distance cross
  sections are calculated in terms of perturbative QCD and LDMEs are obtained by 
  fitting the experimental data. Measured transverse momentum distributions of 
  $\chi_{\rm c}$, $\psi$(2S) and J/$\psi$ in p +{$\bar {\rm p}$} collisions at $\sqrt{s}=$ 1.8, 1.96 TeV 
  and in p+p collisions at $\sqrt{s}=$  7, 8 and 13 TeV are used to constrain LDMEs. The feed-down 
  contribution to each state from the higher states are taken into account.
  The formalism provides a very good description of the data in a wide energy range. 
  The values of LDMEs are used to predict the charmonia cross sections in p+p collisions 
  at 13 and 5 TeV in kinematic bins relevant for the LHC detectors. 
\end{abstract}
\maketitle


\section{Introduction}

 The quarkonia ($Q\bar Q$) have provided useful tools for probing both 
perturbative and nonperturbative aspects of Quantum Chromodynamics (QCD) 
eversince the discovery of J/$\psi$ resonance~\cite{Augustin:1974xw,Aubert:1974js}. 
 The Quarkonia states are qualitatively different from most other hadrons since 
the velocity $v$ of the heavy constituents is small allowing a 
non-relativistic treatment of bound states. 
{\color{black}
 The quarkonia yields are modified in the heavy ion collision due to QGP
  and cold nuclear matter effects which has been demonstrated for J/$\psi$ and $\Upsilon$ 
  in PbPb collisions~\cite{Braun-Munzinger:2015hba,Kumar:2014kfa,Mocsy:2013syh}. The ratios of excited to ground state quarkonia 
  yields are considered as better probes of QGP since the cold matter effects, 
  which are similar for the ground and excited states, are expected to cancel in the 
  ratio. At the LHC, the production of charmonium (J/$\psi$, $\psi$(2S)) 
  and bottomonium ($\Upsilon$(1S),$\Upsilon$(2S),$\Upsilon$(3S) ) states has been studied in 
  PbPb collisions at $\sqrt{s_{NN}} = 2.76$~TeV and $\sqrt{s_{NN}} = 5.02$~TeV
  ~\cite{Chatrchyan:2012lxa,Khachatryan:2014bva,Khachatryan:2016ypw,Sirunyan:2016znt,Khachatryan:2016xxp,Abelev:2013ila}
  affirming the importance of quarkonia measurements in heavy ion collisions.
}
  The heavy quarks due to their high mass 
($m_{\rm c}\,\sim$ 1.6 GeV/c$^2$, $m_{\rm b}\,\sim$ 4.5 GeV/c$^2$), 
are produced in initial partonic collisions with sufficiently high momentum 
transfers. Thus the heavy quark production can be treated 
perturbatively~\cite{Nason:1987xz,Nason:1989zy}.
  The formation of quarkonia out of the two heavy quarks is a nonperturbative 
process and is treated in terms of different 
models~\cite{Bodwin:1994jh,Brambilla:2010cs,Brambilla:2014jmp}. 
  Most notable models for quarkonia production are the color-singlet
model (CSM), the color-evaporation model (CEM), the non-relativistic QCD
(NRQCD) factorization approach, and the fragmentation-function approach.

  In the CSM \cite{Einhorn:1975ua,Ellis:1976fj,Carlson:1976cd,Berger:1980ni},
it is assumed that the $Q\bar Q$ pair that evolves into
the quarkonium is in a color-singlet state and has the same spin
and angular-momentum as the quarkonium. 
 The production rate of quarkonium state is related to 
the absolute values of the color-singlet $Q\bar Q$ wave function and 
its derivatives, evaluated at zero $Q\bar Q$ separation. These quantities 
can be extracted by comparing calculated quarkonium decay
rates in the CSM with the experimental measurements. 
 The CSM was successful in predicting quarkonium production rates at
relatively low energy \cite{Schuler:1994hy} but, at high
energies, very large corrections appear at next-to-leading
order (NLO) and next-to-next-to-leading order (NNLO) in $\alpha_s$
\cite{Artoisenet:2007xi,Campbell:2007ws,Artoisenet:2008fc}.
  The NRQCD factorization approach comprises the color-singlet model, 
but also includes color-octet states.
    In the CEM~\cite{Fritzsch:1977ay,Amundson:1995em,Amundson:1996qr}, it
is assumed that the produced $Q\bar Q$ pair evolves into a quarkonium
if its invariant mass is less than the threshold for producing a 
pair of open-flavor heavy mesons. 
 The nonperturbative probability for the $Q\bar Q$ pair to evolve into 
a quarkonium state is fixed by comparison with the measured production
cross section of that quarkonium state.
 The CEM calculations provide good descriptions of the CDF data for J/$\psi$,
$\psi$(2S), and $\chi_{\rm c}$ production at $\sqrt{s}=1.8$~TeV
\cite{Amundson:1996qr} but it fails to predict the quarkonium 
polarization.

  In the NRQCD factorization approach~\cite{Bodwin:1994jh},
the probability for a $Q\bar Q$ pair to evolve into a quarkonium is expressed
as matrix elements of NRQCD operators in terms of the heavy-quark velocity 
$v$ in the limit $v\,\ll\,1$. This approach takes into account the complete structure of 
the $Q\bar Q$ Fock space, which is spanned by the state $n\,=\,^{2S+1}L_{J}^{[a]}$ 
with spin $S$, orbital angular momentum $L$, total angular momentum $J$, 
and color multiplicity $a$ = 1 (color-singlet), 8 (color-octet). 
 The $Q\bar Q$ pairs which are produced at short distances in color-octet (CO) states, 
evolve into physical, color-singlet (CS) quarkonia by emitting soft gluons 
nonperturbatively.
 In the limit $v\rightarrow0$, the CSM is recovered in the case of S-wave quarkonia.
 The short distance cross sections can be calculated within the 
framework of perturbative QCD (pQCD). The long distance matrix elements (LDME) 
corresponding to the probability of the $Q\barQ$ state to convert to the quarkonium 
can be estimated by comparison with the experimental measurements. 
  The leading order (LO) NRQCD gives a good description of J/$\psi$ yields at 
Tevatron RHIC and LHC energies~\cite{Beneke:1996yw,Braaten:1999qk,Sharma:2012dy}.
  The NLO corrections to color-singlet J/$\psi$ production have been investigated 
in Refs.~\cite{Campbell:2007ws,Gong:2008sn}.
{\color{black}
 The NLO corrections increase the total color-singlet J/$\psi$ cross section by a 
 factor of two, although at high $p_T$ the corrections can enhance the production by
 two-three orders of magnitude.~\cite{Gong:2008sn}.
 The NLO corrections to J/$\psi$ production via S-wave 
 color octet (CO) states ($^1S_{0}^{[8]}\,^3S_{1}^{[8]}$) are studied in 
 Ref.~\cite{Gong:2008ft} and the corrections to $p_{T}$ distributions of both 
 J/$\psi$ yield and polarization are found to be small.
}
In Refs.~\cite{Ma:2010vd}, 
NLO corrections for $\chi_{cJ}$ hadroproduction are also studied. 
Several NLO calculations are performed to obtain the polarization and yield of
J/$\psi$. The J/$\psi$ polarization presents a rather confusing 
pattern~\cite{Butenschoen:2012px, Butenschoen:2012qr, Chao:2012iv,Gong:2012ug}.
{\color{black}
Authors in Ref.~\cite{Butenschoen:2012qr} extracted leading color-octet LDMEs
through a global fit to experimental data of unpolarized J/$\psi$ production in 
pp, p$\overline{\rm p}$, ep, $\gamma \gamma$, 
and e$^{+}$e$^{-}$ collisions. The extracted LDMEs give excellent description of the 
unpolarized J/$\psi$ yields but fail to reproduce the  polarization measured at 
CDF~\cite{Abulencia:2007us}. In another study~\cite{Chao:2012iv}, it is shown 
that the measured hadroproduction cross sections and the CDF polarization 
measurement~\cite{Abulencia:2007us} can be simultaneously described by NRQCD at NLO.
}
The works of Ref.~\cite{Butenschoen:2010rq,Butenschoen:2011yh} and  Ref.~\cite{Ma:2010yw} present
NLO-NRQCD calculations of J/$\psi$ yields. In both the works, the set of CO LDMEs 
fitted to $p_{T}$ distributions measured at HERA and CDF are used to describe 
the $p_{T}$ distributions from RHIC and the LHC.
The fitted LDMEs of Ref.~\cite{Butenschoen:2010rq} and  Ref.~\cite{Ma:2010yw}
are incompatible with each other. A recent work~\cite{Shao:2014yta} gives 
calculations for both the yields and polarizations of charmonia at the Tevatron 
and the LHC where the LDMEs are obtained by fitting the Tevatron data only.

{\color{black}
  Recently, the LHCb measurements of $\eta_{c}$ production~\cite{Aaij:2014bga} 
  is investigated from different points of views by several groups using  
  NRQCD formalism~\cite{Butenschoen:2014dra,Han:2014jya,Zhang:2014ybe}.   
  Ref.~\cite{Butenschoen:2014dra} considered the $\eta_{ c}$ measurement as a challenge of NRQCD 
  while Ref.~\cite{Han:2014jya} shows that the LHCb measurement results in a very 
  strong constraint on the upper bound of the color-octet LDME of J/$\psi$.
  Refs.~\cite{Zhang:2014ybe} obtains the color-singlet LDME for 
  $\eta_c$ by fitting the experiment data to get good description of $\eta_c$ production.
  The prompt double heavy quarkonium production should be a more sensitive testing 
  ground for NRQCD factorization. The experiments at LHC recently published 
  the measurement of double J/$\psi$ production in proton-proton collision at $ \sqrt{s} $ 
  = 7, 8 and 13 TeV~\cite{Aaij:2011yc,Khachatryan:2014iia,Aaboud:2016fzt,Aaij:2016bqq}. 
  Full NLO calculations including all color singlet and color octet contributions 
  for this process in the NRQCD framework are not fully established yet. 
  Authors in Ref.~\cite{He:2015qya} showed that the LO calculations of the prompt double J/$\psi$ 
  production by NRQCD formalism describes the data only qualitatively.
  Authors in Ref.~\cite{Sun:2014gca} present the NLO calculations for the color-singlet channel
  which describe the measured LHCb cross section reasonably well, but fail to reproduce the CMS 
  measurements. The complicated situation suggests that, further study and phenomenological test of NRQCD is still an 
  urgent task.
}

 With the LHC running for several years we now have very high quality quarkonia 
production data in several kinematic regions up to very high transverse momentum 
which could be used to constrain the LDMEs. In this paper, we use CDF data
\cite{Abe:1997yz,Abe:1997jz,Acosta:2004yw,Abulencia:2007bra} along with new LHC data 
\cite{Chatrchyan:2011kc,Khachatryan:2015rra,Chatrchyan:2012ub,Aad:2015duc,ATLAS:2014ala,
Aaij:2012ag,Aaij:2011jh,Aaij:2015rla,Aaij:2013dja} to constrain 
the LDMEs. 
 The feed-down contribution to each state from the higher states are taken into account.
These new LDMEs are then used to predict the J/$\psi$ and $\psi$(2S)
cross-section at 13 TeV and 5 TeV for the kinematical bins relevant to LHC detectors.

 The NLO calculations are still evolving and thus we use LO calculations in 
this work. The values of fitted LDMEs with LO formulations are always useful 
for straightforward predictions of quarkonia cross section and for the 
purpose of a comparison with those obtained using NLO formulations.
{\color{black}
We have given an estimate of uncertainties in the LDMEs due to enhancement of 
color-singlet J/$\psi$ cross-section by a factor of three expected from NLO 
corrections.
}

\section{Quarkonia Production in p$+$p collisions}
\label{section:ppProduction}
 The NRQCD formalism provides a theoretical framework for studying the heavy 
quarkonium production. The dominant processes in the production of heavy 
mesons $\psi$ are $g+q\rightarrow \psi+q$, $q+\bar{q}\rightarrow \psi+g$ and 
$g+g\rightarrow \psi+g$. We represent these processes by $a+b\rightarrow \psi+X$, 
where $a$ and $b$ are the light incident partons. The invariant cross-section 
for the production of a heavy meson $\psi$ can be written in a factorized form as 
\begin{equation}
    E\frac{d^{3}\sigma^{\psi}}{d^{3}p} = \sum_{a,b}\int \int dx_a\,dx_b \, 
    G_{a/p}(x_a,\mu_{F}^{2}) \, G_{b/p}(x_b,\mu_{F}^{2}) \, \frac{\hat s}{\pi}\frac{d\sigma}{d\hat t}
    \times \delta(\hat s + \hat t + \hat u -M^{2}), 
\label{eqn:cross}
\end{equation}
where $G_{a/p}(G_{b/p})$ is the distribution function (PDF) of the incoming parton 
$a(b)$ in the incident proton, which depends on the momentum fraction $x_a(x_b)$
and the factorization scale $\mu_F$. The parton level  Mandelstam variables 
${\hat s}$, ${\hat t}$, and ${\hat u}$
can be expressed in terms of $x_a$, $x_b$ as 
\begin{equation}
\begin{split}
\cs = \,& x_{a}\,x_{b}\,s \\
\ct = \,& M^{2} - x_{a}\,\sqrt{s}\,m_{T}\,e^{-y}\\
\cu = \,& M^{2} - x_{b}\,\sqrt{s}\,m_{T}\,e^{y} ,
\end{split}  
\end{equation}
where $\sqrt{s}$ being the total energy in the centre-of-mass, $y$ is the rapidity 
and $p_{T}$ is the transverse momentum of the $Q\bar Q$ pair. The mass of heavy
meson is represented by $M$ and $m_{T}$ is the transverse mass 
defined as $m_{T}^{2} = p_{T}^{2} + M^{2}$. Writing 
down $ \hat s + \hat t + \hat u -M^{2} = 0$ and solving for $x_{b}$, we obtain
\begin{equation}
x_b = \frac{1}{\sqrt{s}}\frac{x_a\,\sqrt{s}\,m_T\,e^{-y}-M^2}{x_a\,\sqrt{s}-m_T\,e^y}.
\end{equation}
 The double differential cross-section upon $p_{T}$ and $y$ then is obtained as
\begin{equation}
\frac{{d^{2}\sigma}^{\psi}}{dp_T\,dy} = \sum_{a,b}\int_{x_{a}^{min}}^{1} dx_a\, 
           G_{a/A}(x_a,\,\mu^{2}_{F})\, G_{b/B}(x_b,\,\mu^{2}_{F})\times 
            2p_T \frac{x_a\,x_b}{x_a-\frac{m_T}{\sqrt{s}}e^y}\frac{d\sigma}{d\hat t},
\end{equation}
where the minimum value of $x_a$ is given by
\begin{equation}
x_{a\rm min} = \frac{1}{\sqrt{s}}\frac{\sqrt{s}\,m_T\,e^{y}-M^2}{\sqrt{s}-m_T\,e^{-y}}.
\end{equation}

 The parton level cross-section $d\sigma/d\hat{t}$ is defined as~\cite{Bodwin:1994jh}
\begin{equation}
\frac{d\sigma}{d\hat t} = \frac{d\sigma}{d\hat t}(ab\rightarrow Q\overline{Q}(^{2S+1}L_{J})+X)
               \, M_{L}(Q\overline{Q}(^{2S+1}L_{J})\rightarrow\psi).
\end{equation}
  The short distance contribution 
$d\sigma/d\hat t (ab\rightarrow Q\overline{Q}(^{2S+1}L_{J})+X)$ 
corresponds to the production of a $Q\barQ$ pair in a particular
color and spin configuration can be calculated within the framework of 
perturbative QCD (pQCD). The long distance matrix elements (LDME) 
$M_{L}(Q\overline{Q}(^{2S+1}L_{J})\rightarrow\psi)$ corresponds to the 
probability of the $Q\barQ$ state to convert to the quarkonium wavefunction
and can be estimated by comparison with experimental measurements. 
 The short distance invariant differential cross-section is given by
\begin{equation}
  \frac{d\sigma}{d\hat t}(ab\rightarrow Q\overline{Q}(^{2S+1}L_{J})+X) 
                = \frac{|\mathcal{M}|^2}{16\pi{\hat s}^2},
\end{equation}
where $|\mathcal{M}|^2$ is the Feynman squared amplitude. We use the expressions for the 
short distance CS cross-sections given in 
Refs.~\cite{Baier:1983va,Humpert:1986cy,Gastmans:1987be} and the CO 
cross-sections given in Refs.~\cite{Cho:1995vh,Cho:1995ce,Braaten:2000cm}. 
  The CTEQ6M~\cite{Lai:2010vv} parametrization is used for parton 
distribution functions. 

   The LDMEs scale with a definite power of the relative velocity $v$ of the 
heavy quarks inside $Q\bar Q$ bound states. In the limit $v<<1$, the production of 
quarkonium is based on the $^3S_1^{[1]}$ and $^3P_J^{[1]}$ ($J$ = 0,1,2) CS states 
and $^1S_0^{[8]}$, $^3S_1^{[8]}$ and $^3P_J^{[8]}$ CO states. The differential 
cross section for the direct production of $J/\psi$ can be written as the 
sum of these contributions,


\begin{equation}
\begin{split}
d\sigma(J/\psi) &=d\sigma(Q\barQ([^3S_1]_{1}))\,M_{L}(Q\barQ([^3S_1]_{1})\rightarrow J/\psi)\\ 
                &+d\sigma(Q\barQ([^1S_0]_{8}))\,M_{L}(Q\barQ([^1S_0]_{8})\rightarrow J/\psi)\\ 
                &+d\sigma(Q\barQ([^3S_1]_{8}))\,M_{L}(Q\barQ([^3S_1]_{8})\rightarrow J/\psi)\\ 
                &+d\sigma(Q\barQ([^3P_0]_{8}))\,M_{L}(Q\barQ([^3P_0]_{8})\rightarrow J/\psi)\\ 
                &+d\sigma(Q\barQ([^3P_1]_{8}))\,M_{L}(Q\barQ([^3P_1]_{8})\rightarrow J/\psi)\\
                &+d\sigma(Q\barQ([^3P_2]_{8}))\,M_{L}(Q\barQ([^3P_2]_{8})\rightarrow J/\psi)\\
                &+ \cdot\cdot\cdot  
\label{eq:dsigmaJ}
\end{split}
\end{equation}

  The dots represent contribution of terms at higher powers of $v$. The 
contributions from the CO matrix elements in Eq.~\ref{eq:dsigmaJ} are suppressed 
by $v^4$ compared to the CS matrix elements.

  For the case of the $p$-wave bound states $\chi_{cJ}$ ($\chi_{c0}$, $\chi_{c1}$ 
and $\chi_{c2}$), the color-singlet state $Q\barQ[^3P_J]_{1}$ and the color-octet 
state $Q\barQ[^3S_1]_{8}$ contribute to the same order in $v$ ($v^{5}$) because of 
the angular momentum barrier for the $p-$wave states, and hence both need to be included. 
The $\chi_{c}$ differential cross section thus can be written as 
\begin{equation}
\begin{split}
d\sigma(\chi_{cJ}) &=d\sigma(Q\barQ([^3P_J]_{1}))\,M_{L}(Q\barQ([^3P_J]_{1})\rightarrow \chi_{cJ})\\ 
                  &+d\sigma(Q\barQ([^3S_1]_{8}))\,M_{L}(Q\barQ([^3S_1]_{8})\rightarrow \chi_{cJ})\\
                  &+\cdot\cdot\cdot  
\label{eq:dsigmachi}
\end{split}
\end{equation}
  The prompt $J/\psi$ production at LHC energies consists of direct $J/\psi$ 
production from the initial parton-parton hard scattering and the feed-down contributions 
to the $J/\psi$ from the decay of heavier charmonium states $\psi(2S)$, 
$\chi_{c0}$, $\chi_{c1}$ and $\chi_{c2}$.  The relevant branching fractions are given 
in the Table~\ref{table:CharmoniaBFs} \cite{Nakamura:2010zzi}. 
The prompt $\psi(2S)$ has no significant feed-down contributions from the higher mass states. 

\begin{table*}[h]
\caption{Relevant branching fractions for charmonia~\cite{Nakamura:2010zzi}}.
\begin{tabular}{c|cccc}
Meson From &to $\chi_{c0}$ &to $\chi_{c1}$ &to $\chi_{c2}$ &to $J/\psi$\\ 
\hline
$\psi(2S)$& 0.0962 & 0.092 & 0.0874 & 0.595   \\
$\chi_{c0}$& &  &  & 0.0116           \\
$\chi_{c1}$& &  &  & 0.344           \\
$\chi_{c2}$& &  &  & 0.195           \\
\hline
\end{tabular}
\label{table:CharmoniaBFs}
\end{table*}

  The expressions and the values for the color-singlet operators can be found 
in~\cite{Cho:1995ce,Cho:1995vh,Eichten:1994gt} which are obtained by solving the 
non-relativistic wavefunctions.
 The CO operators can not be related to the non-relativistic
wavefunctions of $Q\barQ$ since it involves a higher Fock state and thus
measured data is used to constrain them.
  The color-singlet contributions along with their calculated values 
and color-octet contributions to be fitted are written below for the prompt
$J/\psi$.
\begin{enumerate}
\item{Direct contributions
\begin{equation}
\begin{split}
&M_{L}(c\barc([^3S_1]_{1})\rightarrow J/\psi) = 1.2\,{\rm GeV}^{3}\\
&M_{L}(c\barc([^3S_1]_{8})\rightarrow J/\psi) \\
&M_{L}(c\barc([^1S_0]_{8})\rightarrow J/\psi) \\
&M_{L}(c\barc([^3P_0]_{8})\rightarrow J/\psi) \\
&M_{L}(c\barc([^3P_1]_{8})\rightarrow J/\psi)= 3\,M_{L}(c\barc([^3P_0]_{8})\rightarrow J/\psi)~\cite{Cho:1995vh}\\ 
&M_{L}(c\barc([^3P_2]_{8})\rightarrow J/\psi)= 5\,M_{L}(c\barc([^3P_0]_{8})\rightarrow J/\psi)~\cite{Cho:1995vh}
~\label{eq:Mpsi1}
\end{split}
\end{equation}
    }
\item{Feed-down contribution from $\psi(2S)$
\begin{equation}
\begin{split}
&M_{L}(c\barc([^3S_1]_{1})\rightarrow \psi(2S)) = 0.76\,{\rm GeV}^{3}\\
&M_{L}(c\barc([^3S_1]_{8})\rightarrow \psi(2S)) \\
&M_{L}(c\barc([^1S_0]_{8})\rightarrow \psi(2S)) \\
&M_{L}(c\barc([^3P_0]_{8})\rightarrow \psi(2S))\\
&M_{L}(c\barc([^3P_1]_{8})\rightarrow \psi(2S))=3\,M_{L}(c\barc([^3P_0]_{8})\rightarrow \psi(2S))~\cite{Cho:1995vh}\\
&M_{L}(c\barc([^3P_2]_{8})\rightarrow \psi(2S))=5\,M_{L}(c\barc([^3P_0]_{8})\rightarrow \psi(2S))~\cite{Cho:1995vh}
~\label{eq:Mpsi2}
\end{split}
\end{equation}
  }
\item{Feed-down contribution from $\chi_{cJ}$
\begin{equation}
\begin{split}
&M_{L}(c\barc([^3P_0]_{1})\rightarrow \chi_{c0}) = 0.054\,m_{c}^{2}\,{\rm GeV}^{5}\\ 
&M_{L}(c\barc([^3S_1]_{8})\rightarrow \chi_{c0})
~\label{eq:Mchi}
\end{split}
\end{equation}
    }
\end{enumerate}
The mass of the charm quark is taken as $m_{c}=1.6$ GeV. The short distance cross 
sections $d\sigma(Q\barQ([^1S_0]_{8}))$ and $d\sigma(Q\barQ([^3P_J]_{8}))$ have
very similar $p_{T}$ dependence and due to this reason the transverse momentum distribution 
is sensitive only to a linear combination of their LDMEs.
Following the Ref.~\cite{Cho:1995vh,Beneke:1996yw} we fit a linear combination 
\begin{equation}
  \begin{split}
    M_{L}(Q\barQ([^1S_0]_{8},[^3P_0]_{8})\rightarrow \psi)&= \frac{M_{L}(Q\barQ([^1S_0]_{8})\rightarrow \psi)}{3} + \frac{M_{L}(Q\barQ([^3P_0]_{8})\rightarrow \psi)}{m_{c}^{2}} \nonumber
  \end{split}
\end{equation}
in our calculations.


\begin{figure}
\begin{minipage}{1.0\linewidth}
\centering
{\includegraphics[width=0.49\textwidth]{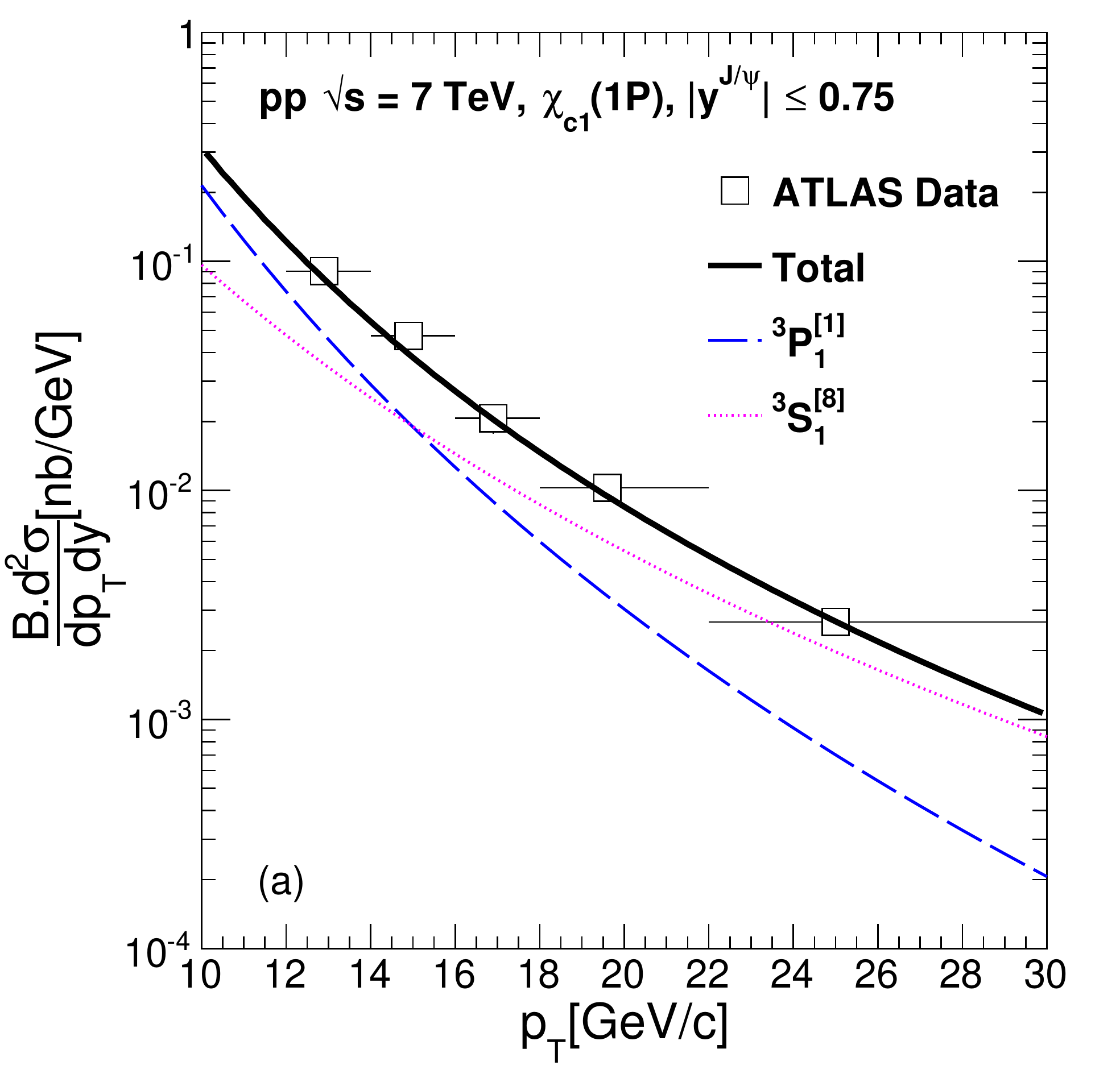}}
{\includegraphics[width=0.49\textwidth]{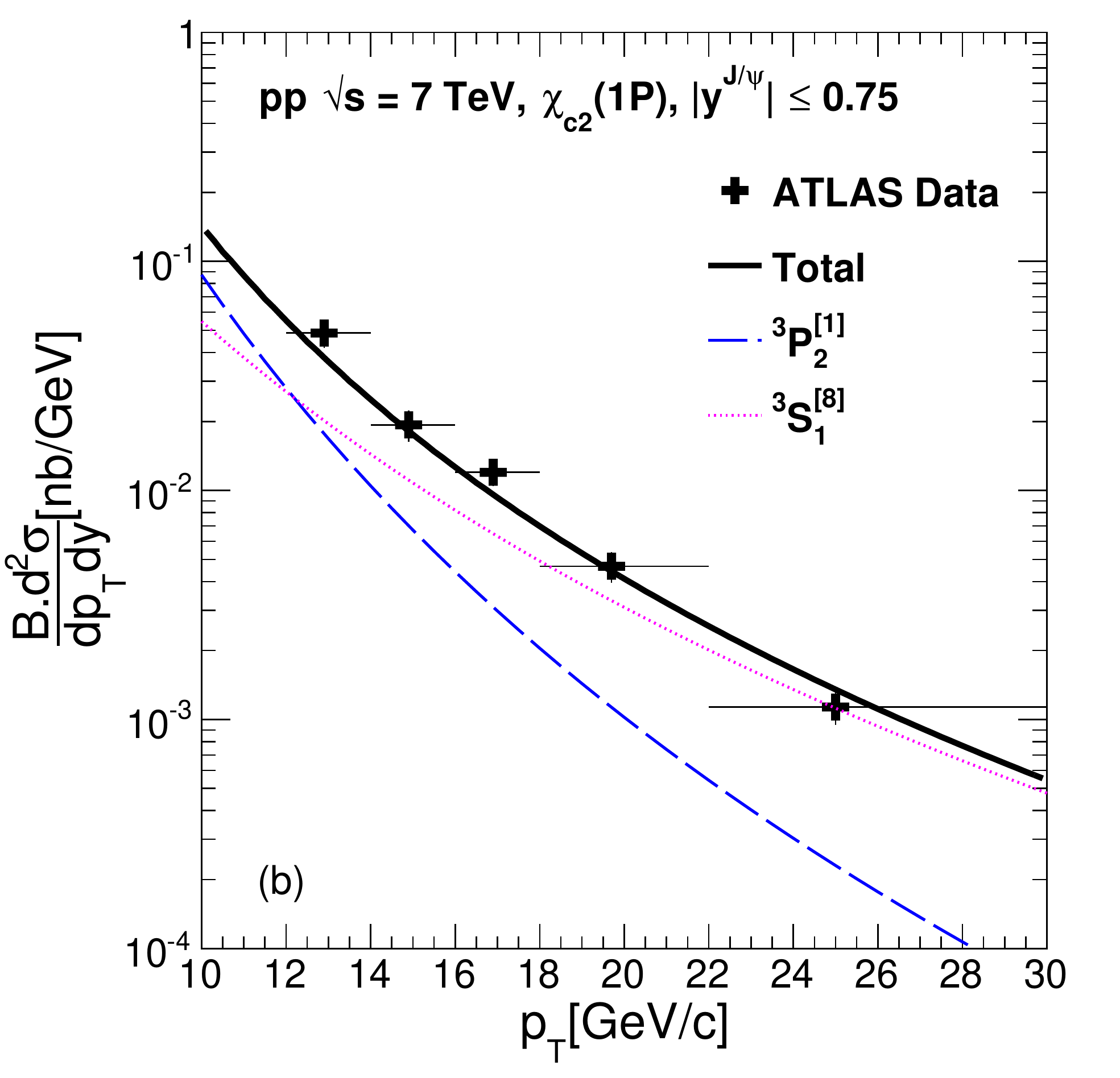}}
\end{minipage}%
\ \\
\centering
\begin{minipage}{0.5\linewidth}
\centering
{\includegraphics[width=1.0\textwidth]{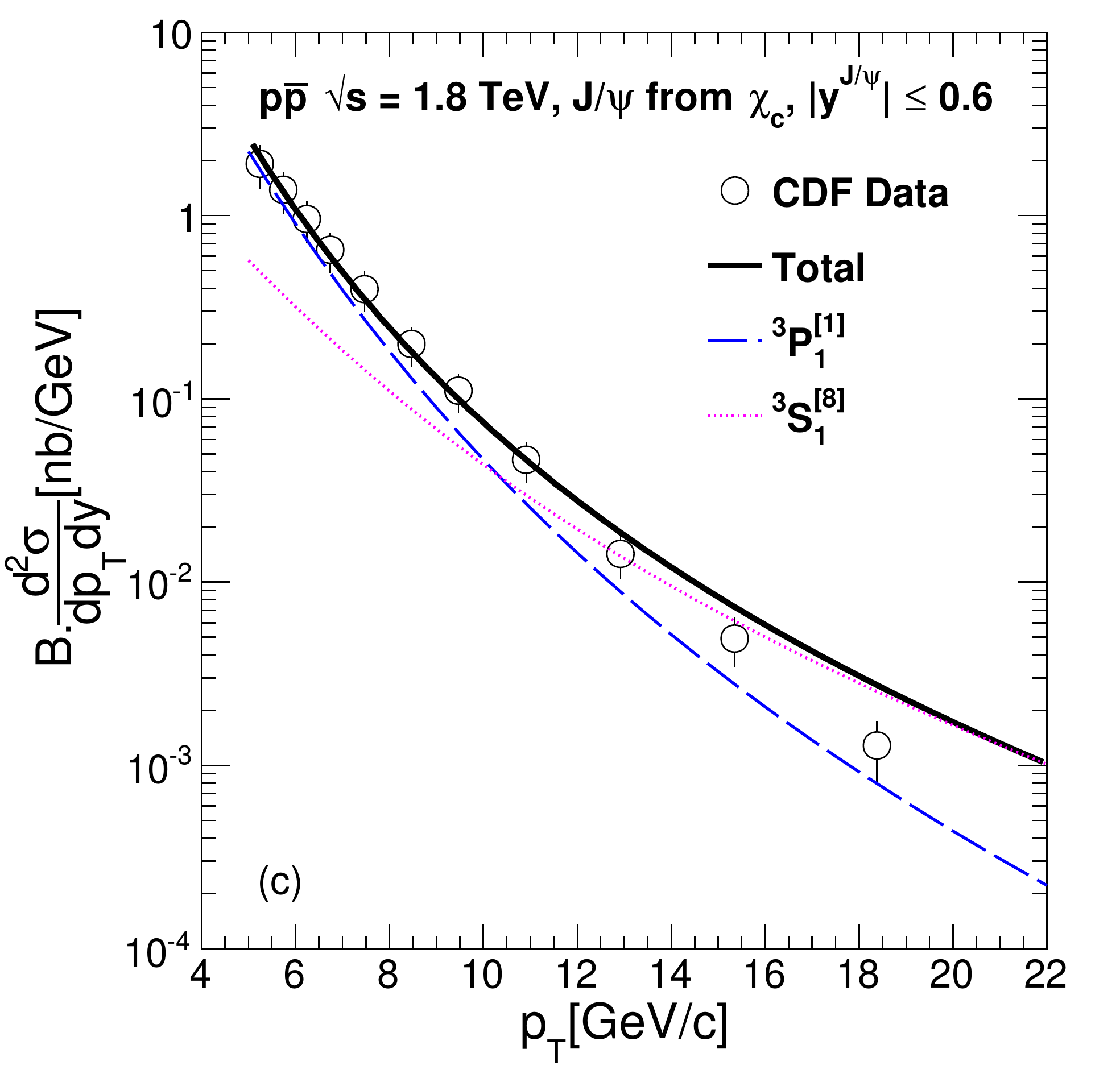}}
\end{minipage}%
\caption{(Color online) The NRQCD calculations of production cross section 
of (a) $\chi_{c1}$, (b) $\chi_{c2}$ in p+p collisions at
$\sqrt{s}$ = 7 TeV and (c) $J/\psi$ from $\chi_{c1}$ and $\chi_{c2}$ 
decays in p+${\bar {\rm p}}$ collisions at $\sqrt{s}$ = 1.8 TeV as 
a function of transverse momentum. The calculations are compared with 
the measured data by ATLAS experiment at LHC~\cite{ATLAS:2014ala}
and measured data by CDF experiment at Tevatron~\cite{Abe:1997yz}. 
The $\chi_{c}$ color octet LDMEs are obtained by fitting this data.}
\label{Fig:LDMEChicATLAS}
\end{figure}

\begin{figure}
\includegraphics[width=0.49\textwidth]{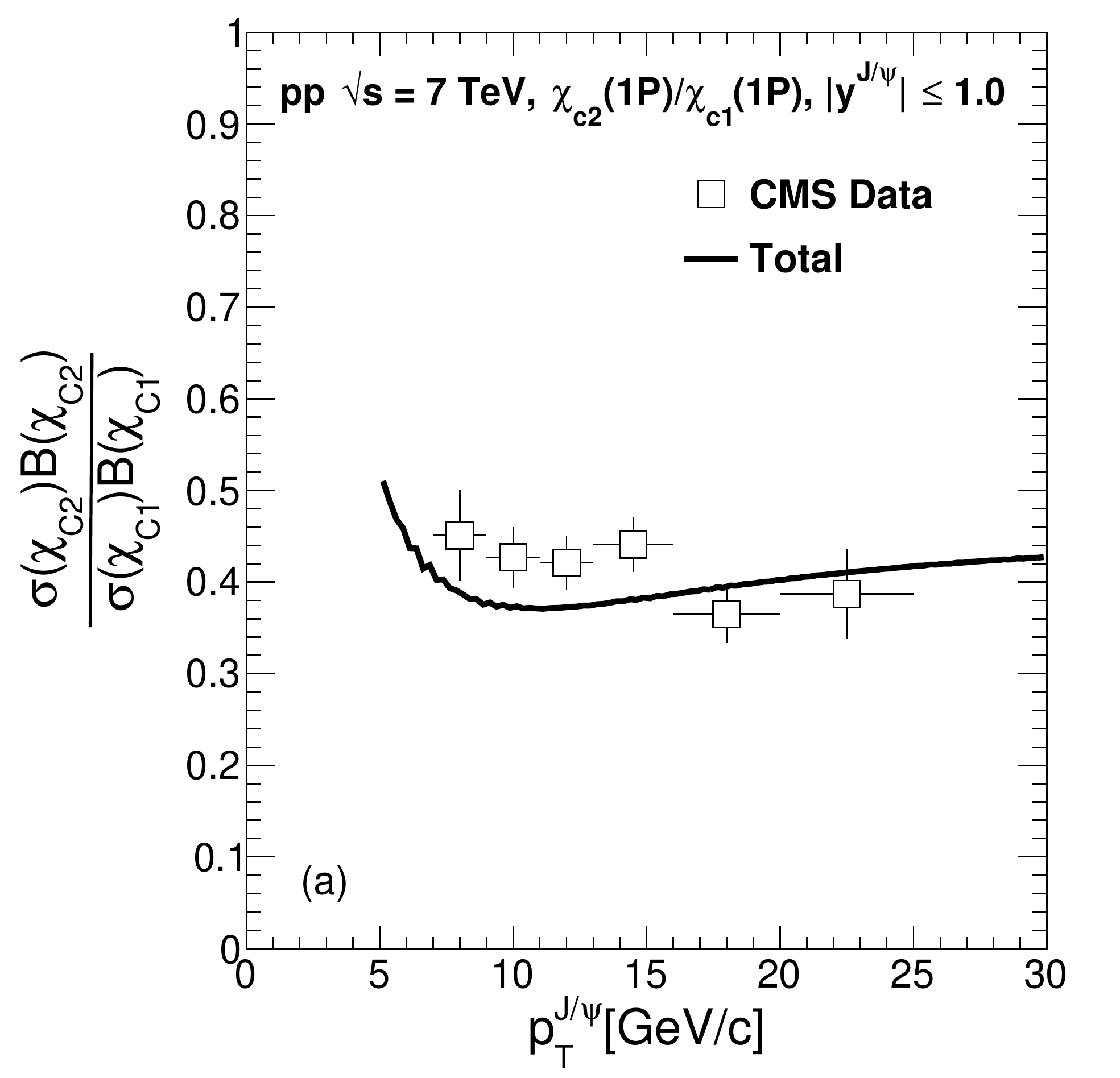}
\includegraphics[width=0.49\textwidth]{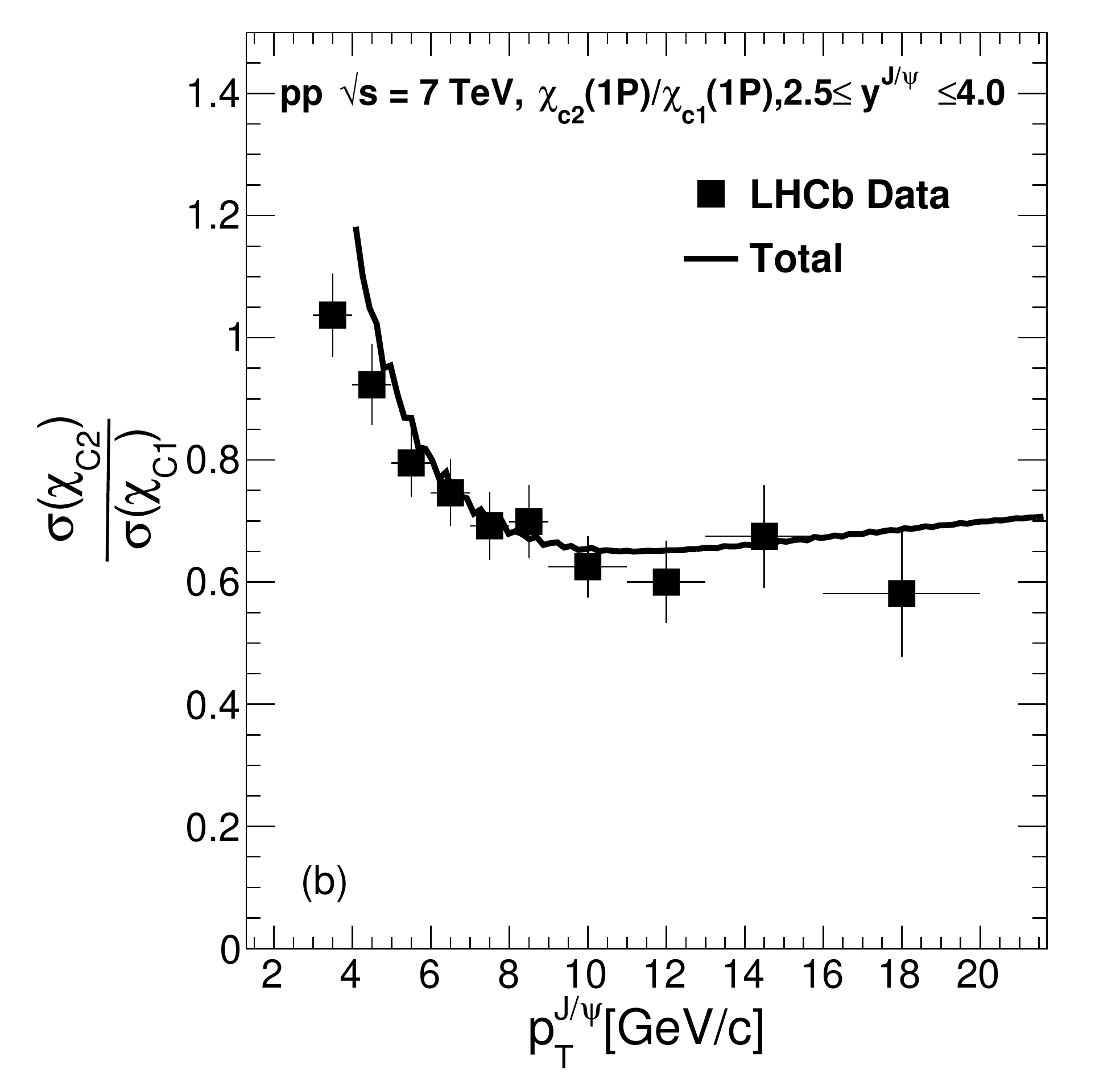}
\caption{(Color online) The NRQCD calculations of production cross section ratios 
of  $\chi_{c2}$ and $\chi_{c1}$ in p+p collisions at
$\sqrt{s}$ = 7 TeV as a function of transverse momentum. 
The calculations are compared with the measured data by
CMS and LHCb experiments at LHC~\cite{Chatrchyan:2012ub,Aaij:2013dja}. The $\chi_{c}$
color octet LDMEs are obtained by fitting this data. 
}
\label{Fig:LDMEChicCMS_LHCb}
\end{figure}

\begin{figure}
\includegraphics[width=0.49\textwidth]{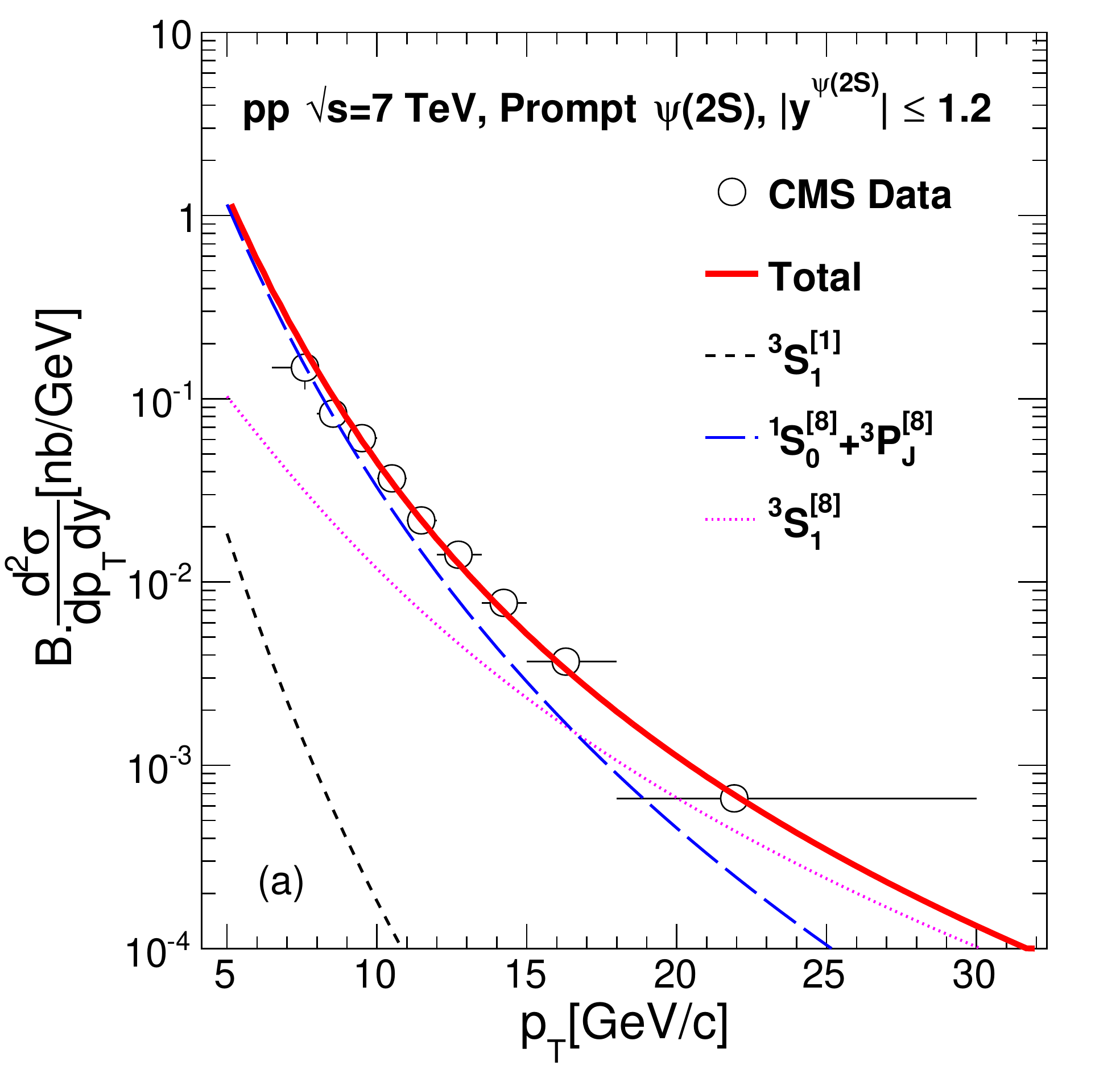}
\includegraphics[width=0.49\textwidth]{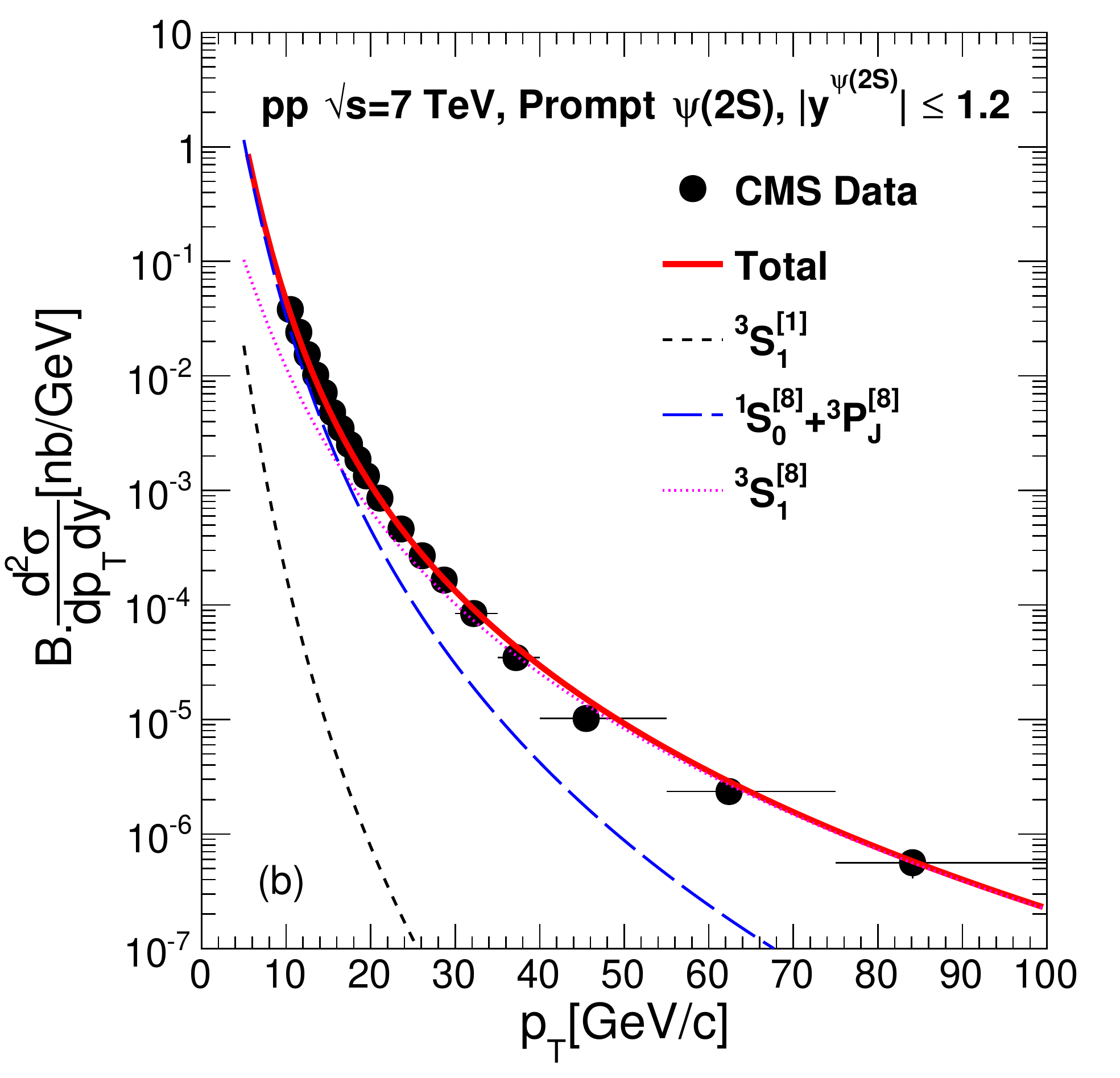}
\includegraphics[width=0.49\textwidth]{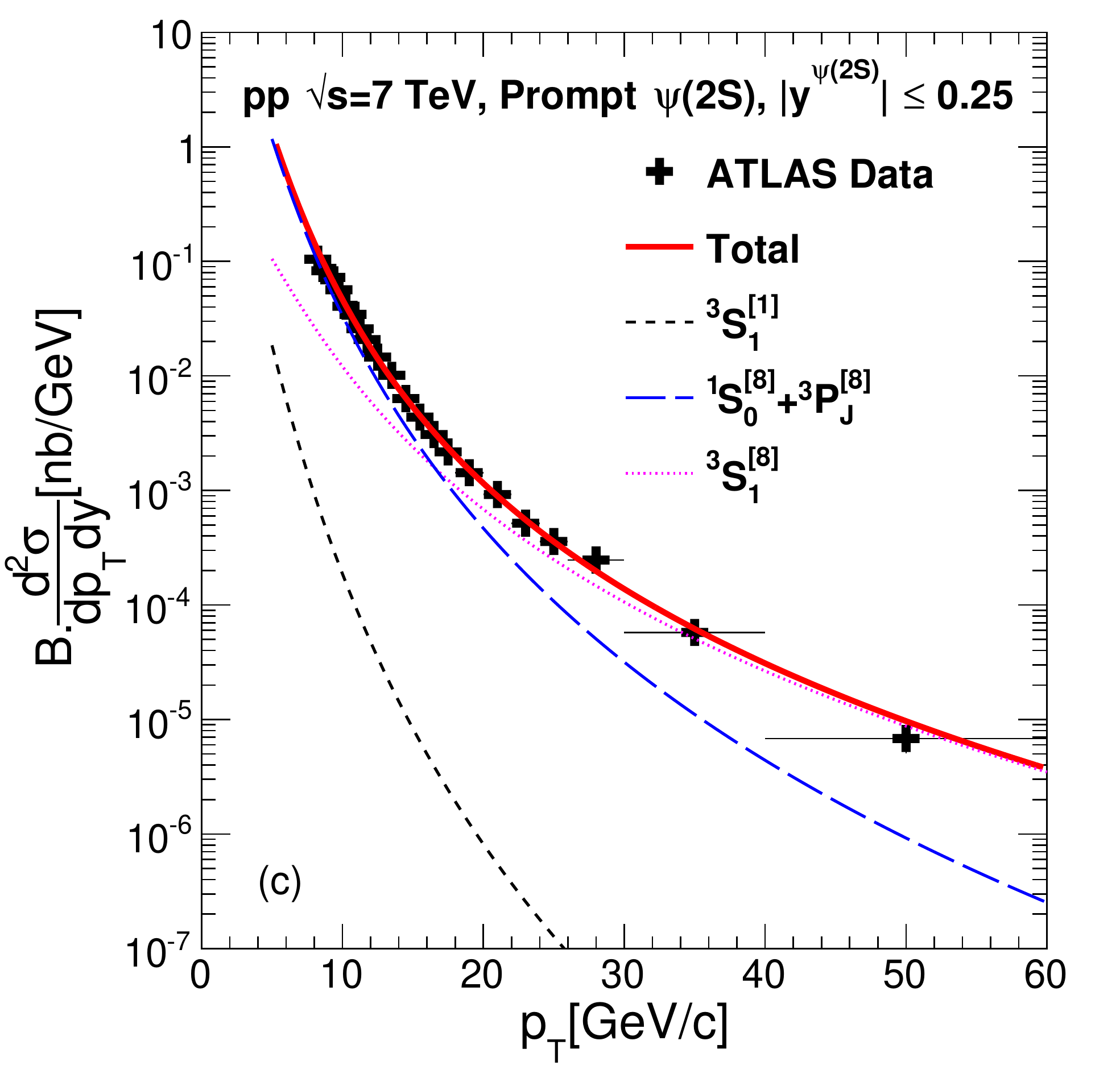}
\includegraphics[width=0.49\textwidth]{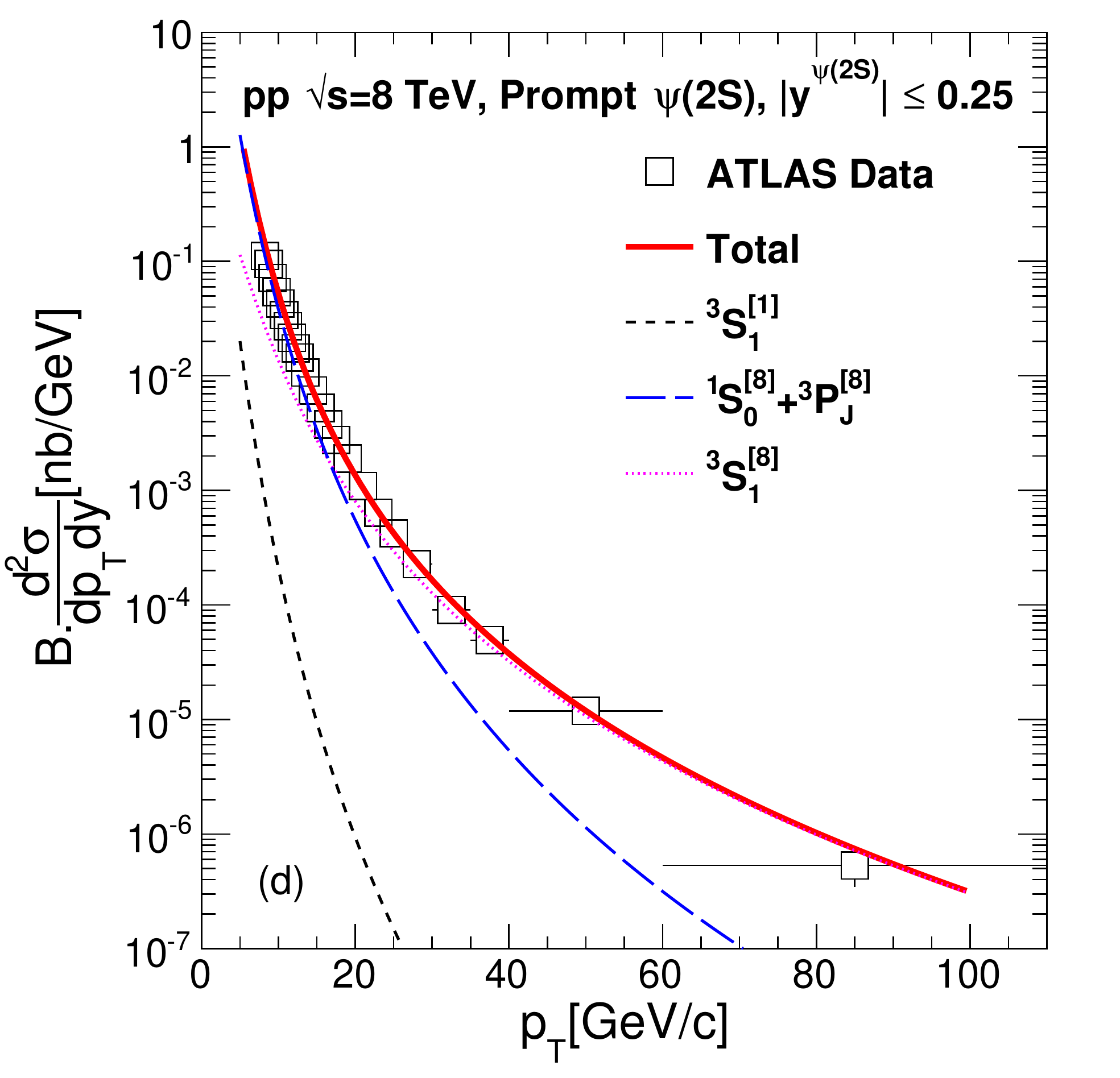}
\caption{(Color online) 
  The NRQCD calculations of production cross section of $\psi$(2S) in p+p collisions 
as a function of transverse momentum compared with the measured data at LHC 
 (a) CMS data at $\sqrt{s}$ = 7 TeV~\cite{Chatrchyan:2011kc} 
 (b) CMS data at $\sqrt{s}$ = 7 TeV~\cite{Khachatryan:2015rra} 
 (c) ATLAS data at $\sqrt{s}$ = 7 TeV and
 (d) ATLAS data at $\sqrt{s}$ = 8 TeV~\cite{Aad:2015duc}. 
 The LDMEs are obtained by a combined fit of the LHC and Tevatron data.
}
\label{Fig:LDMEPsi2S}
\end{figure}


\begin{figure}
\begin{minipage}{1.0\linewidth}
\centering
{\includegraphics[width=0.49\textwidth]{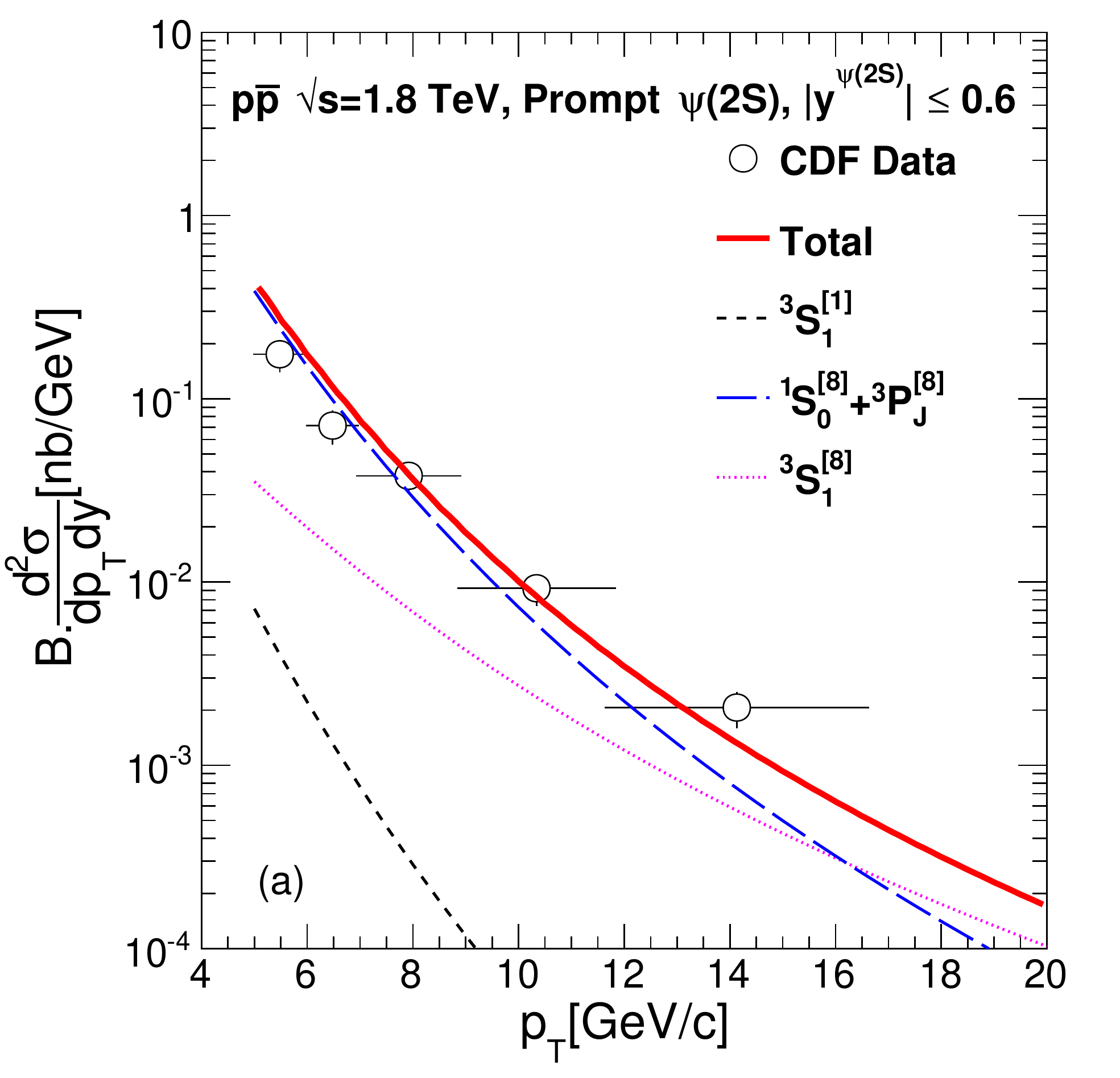}}
{\includegraphics[width=0.49\textwidth]{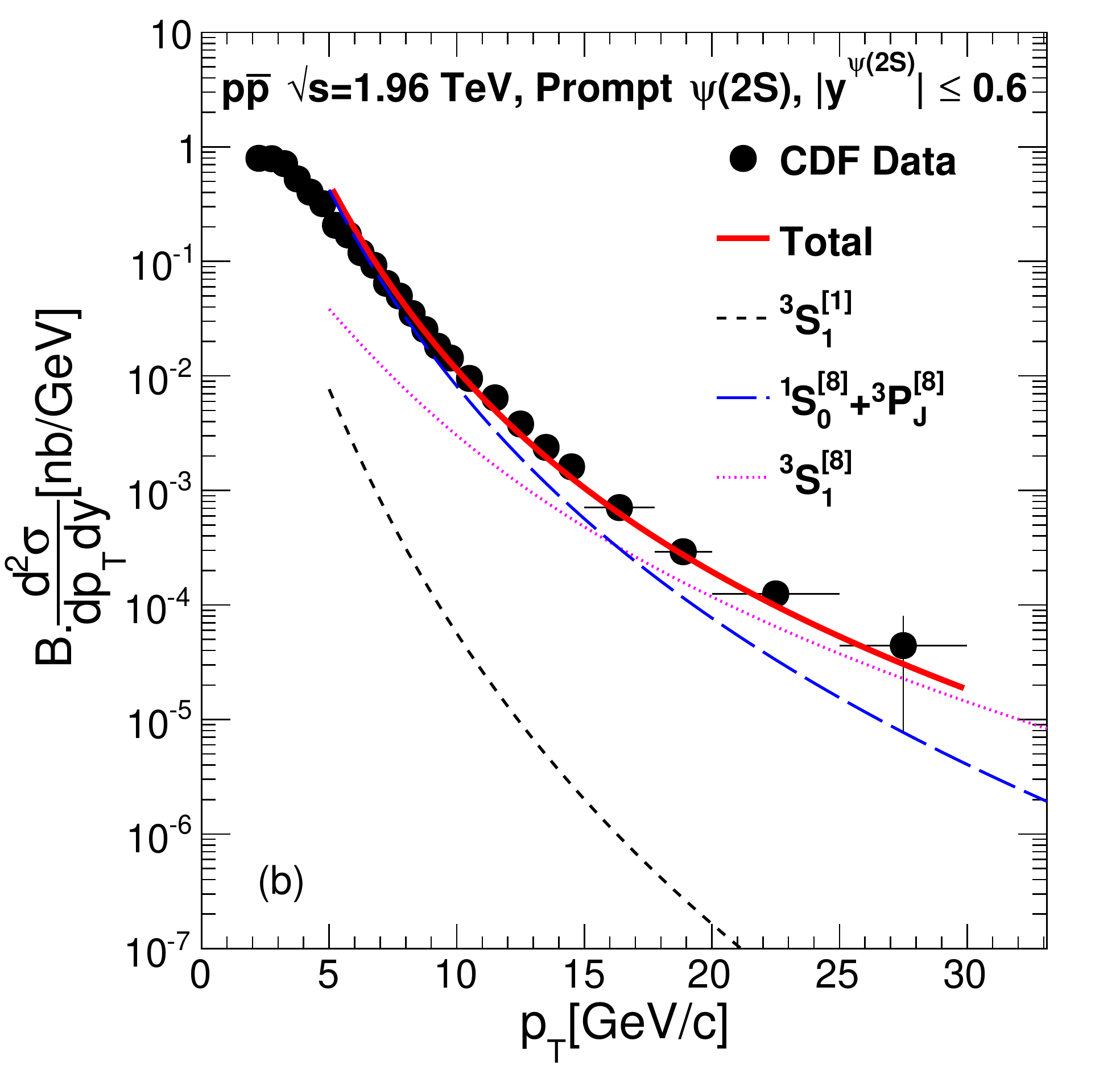}}
\end{minipage}%
\ \\
\centering
\begin{minipage}{0.5\linewidth}
\centering
{\includegraphics[width=1.0\textwidth]{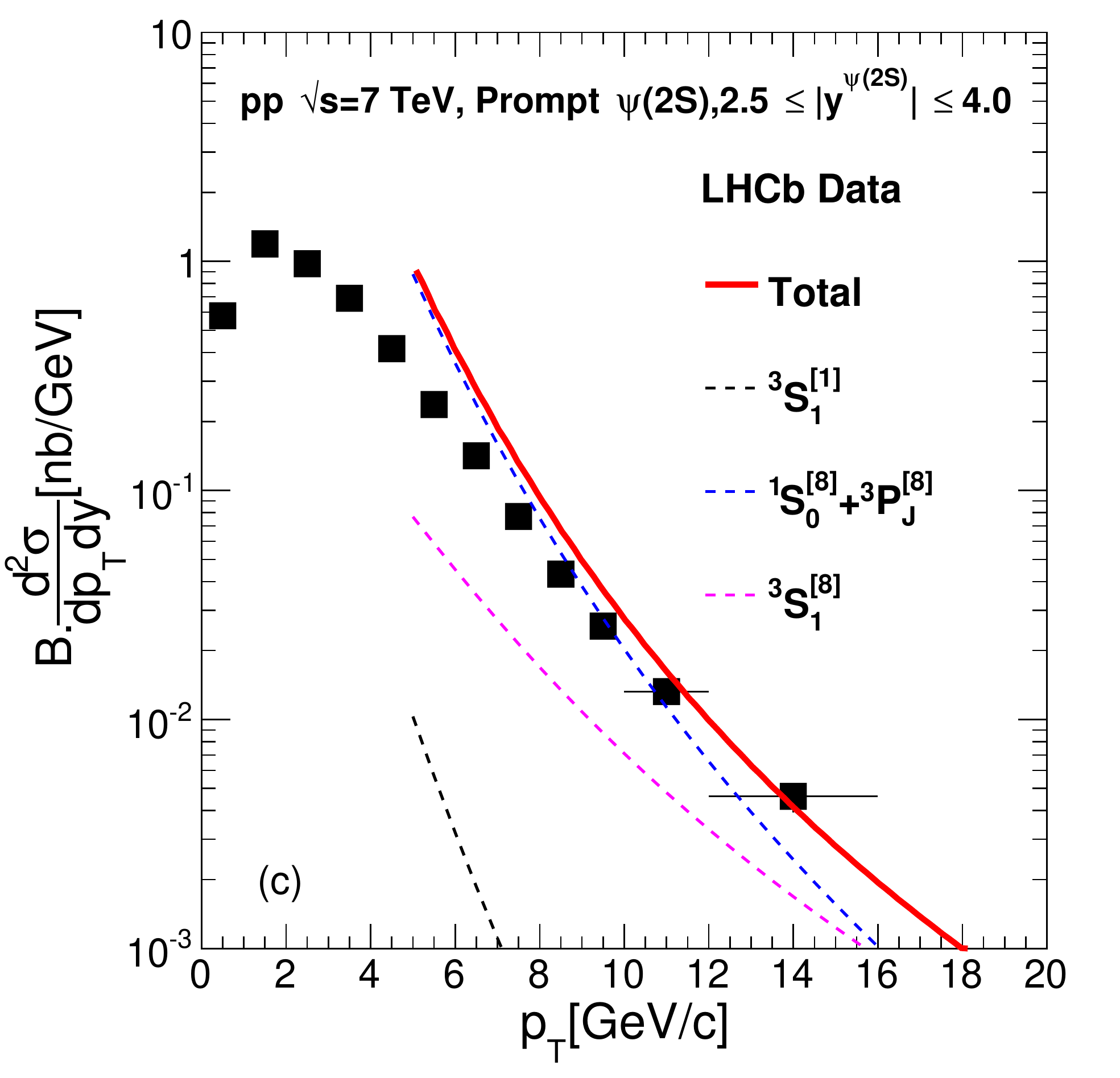}}
\end{minipage}%
\caption{(Color online) The NRQCD calculations of production cross section 
of $\psi$(2S) in p+${\bar {\rm p}}$ and p+p collisions as a function of transverse 
momentum compared with the measured  data 
(a) CDF data at $\sqrt{s}$ = 1.8 TeV~\cite{Abe:1997jz}, 
(b) CDF data at $\sqrt{s}$ = 1.96 TeV~\cite{Acosta:2004yw} and 
(c) LHCb data at $\sqrt{s}$ = 7 TeV~\cite{Aaij:2012ag}.
 The LDMEs are obtained by a combined fit of the Tevatron and
LHC data.}
\label{Fig:LDMEPsi2SCDF}
\end{figure}

\begin{figure}
\includegraphics[width=0.49\textwidth]{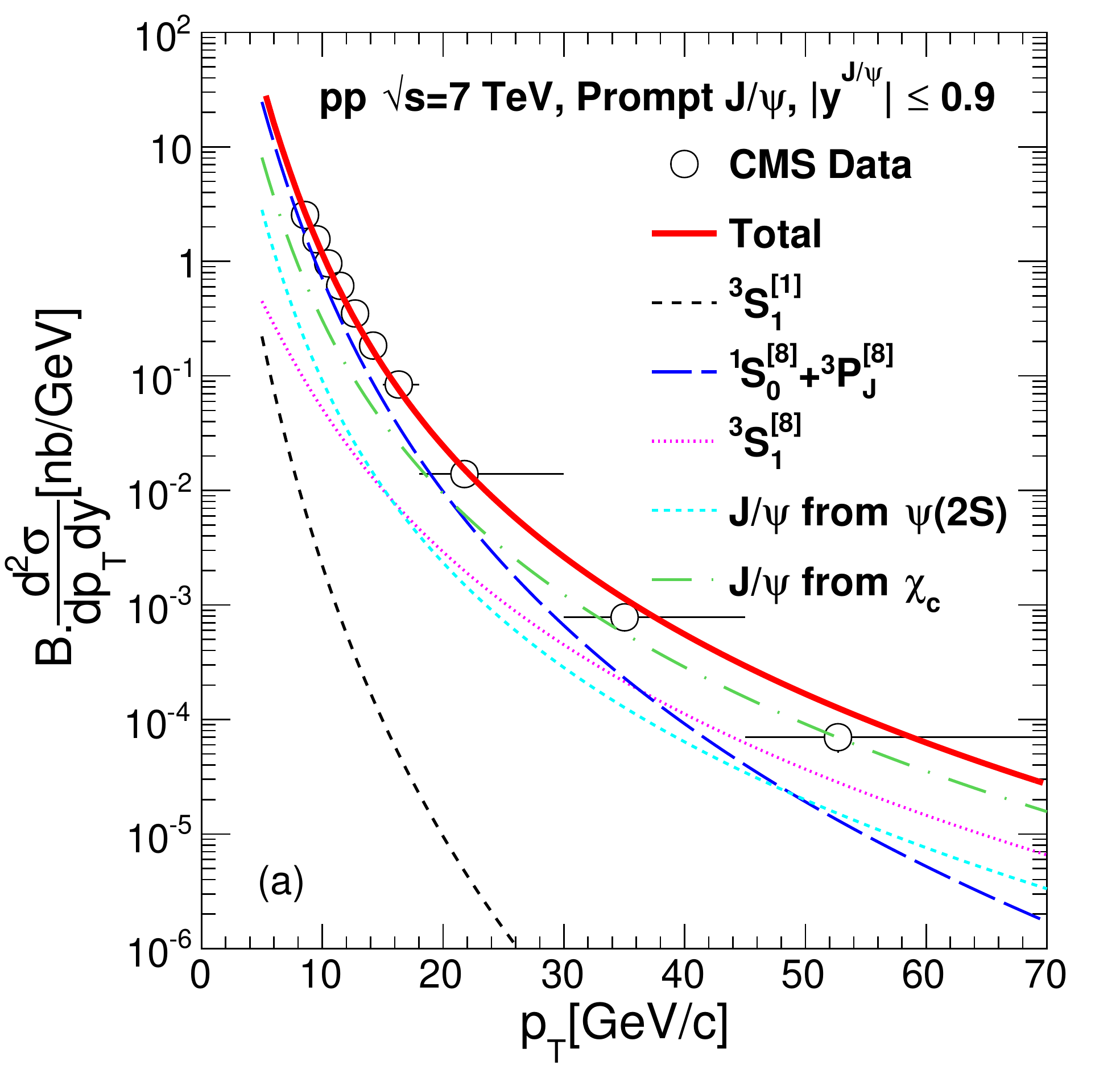}
\includegraphics[width=0.49\textwidth]{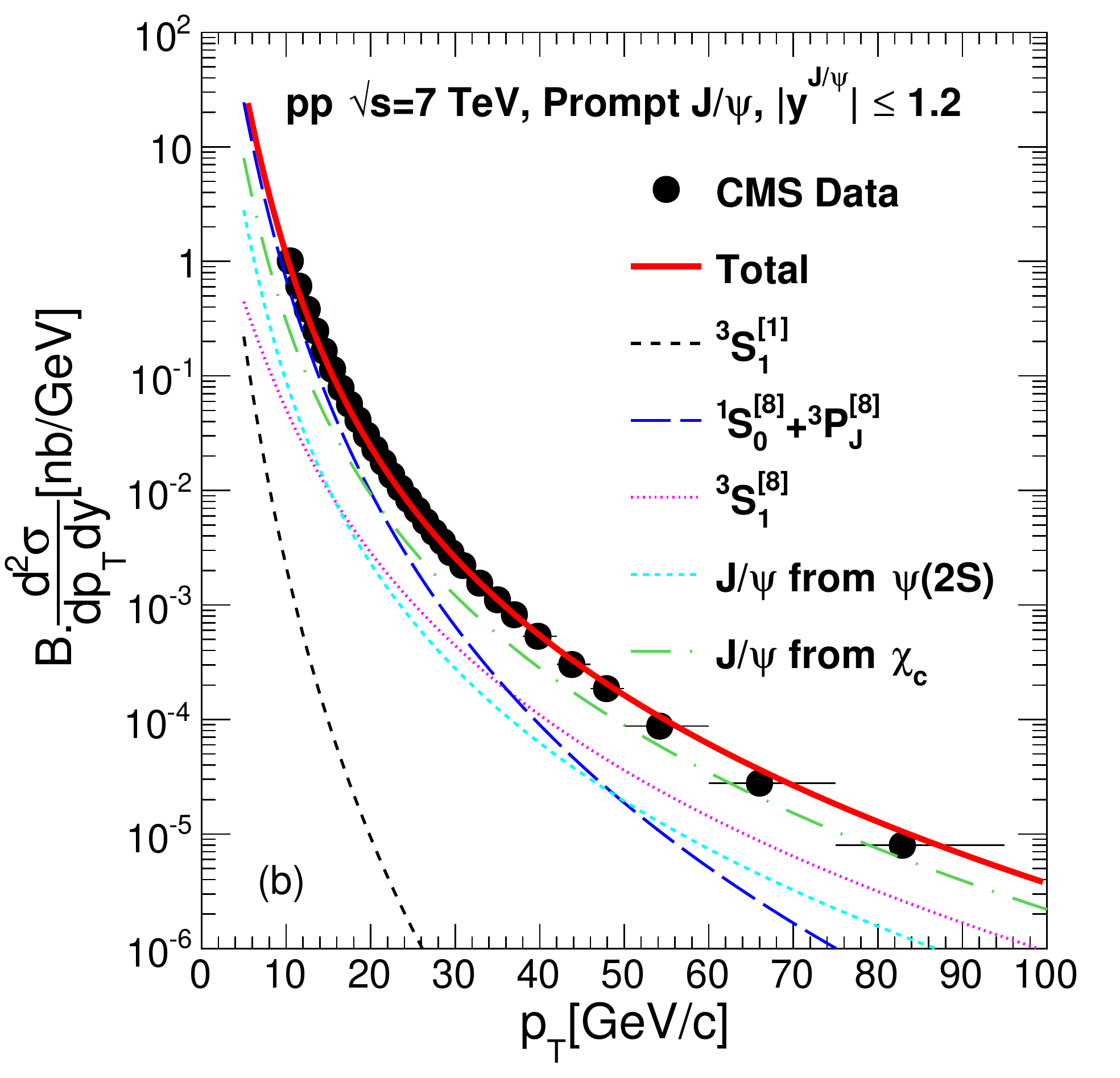}
\includegraphics[width=0.49\textwidth]{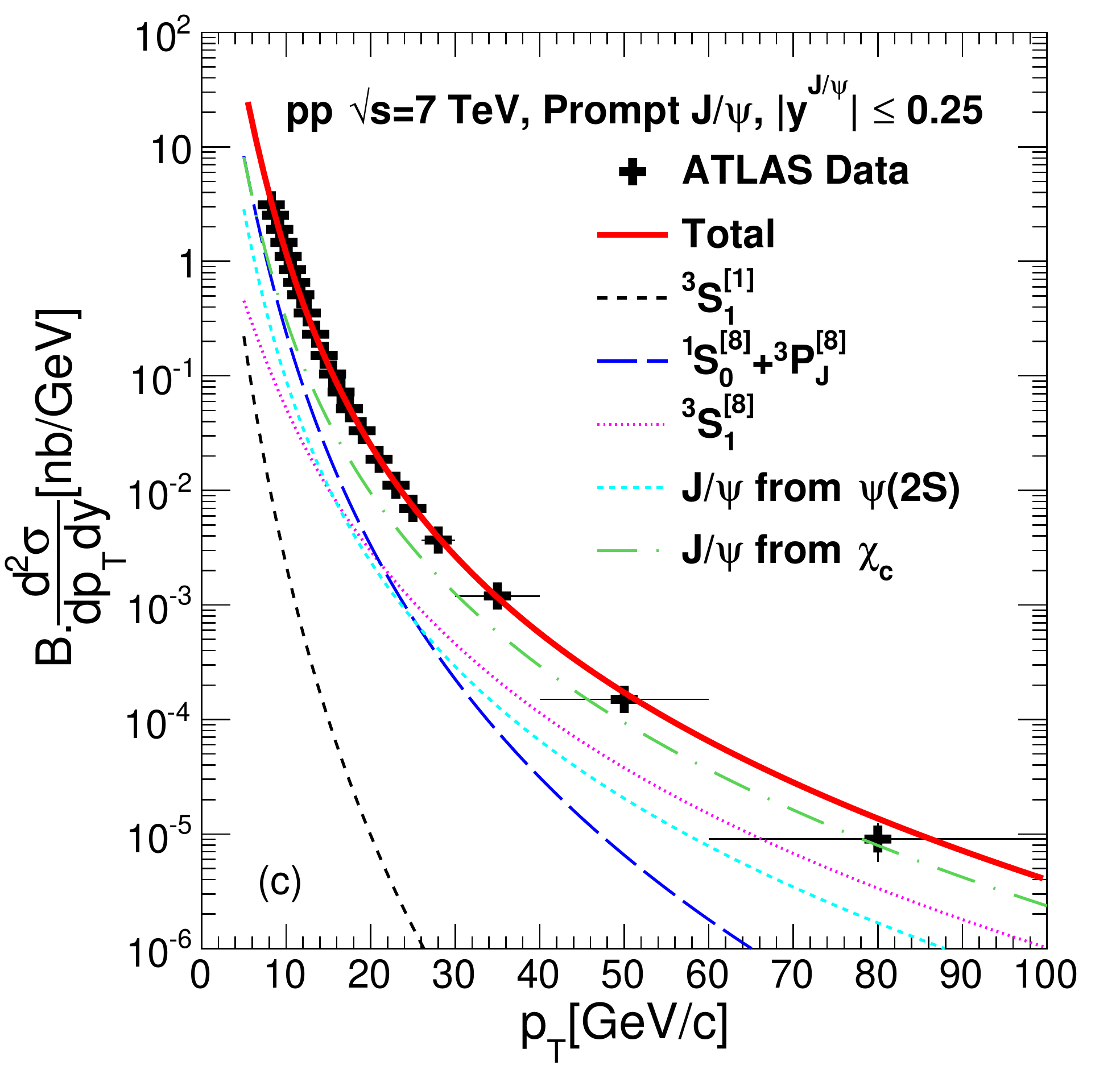}
\includegraphics[width=0.49\textwidth]{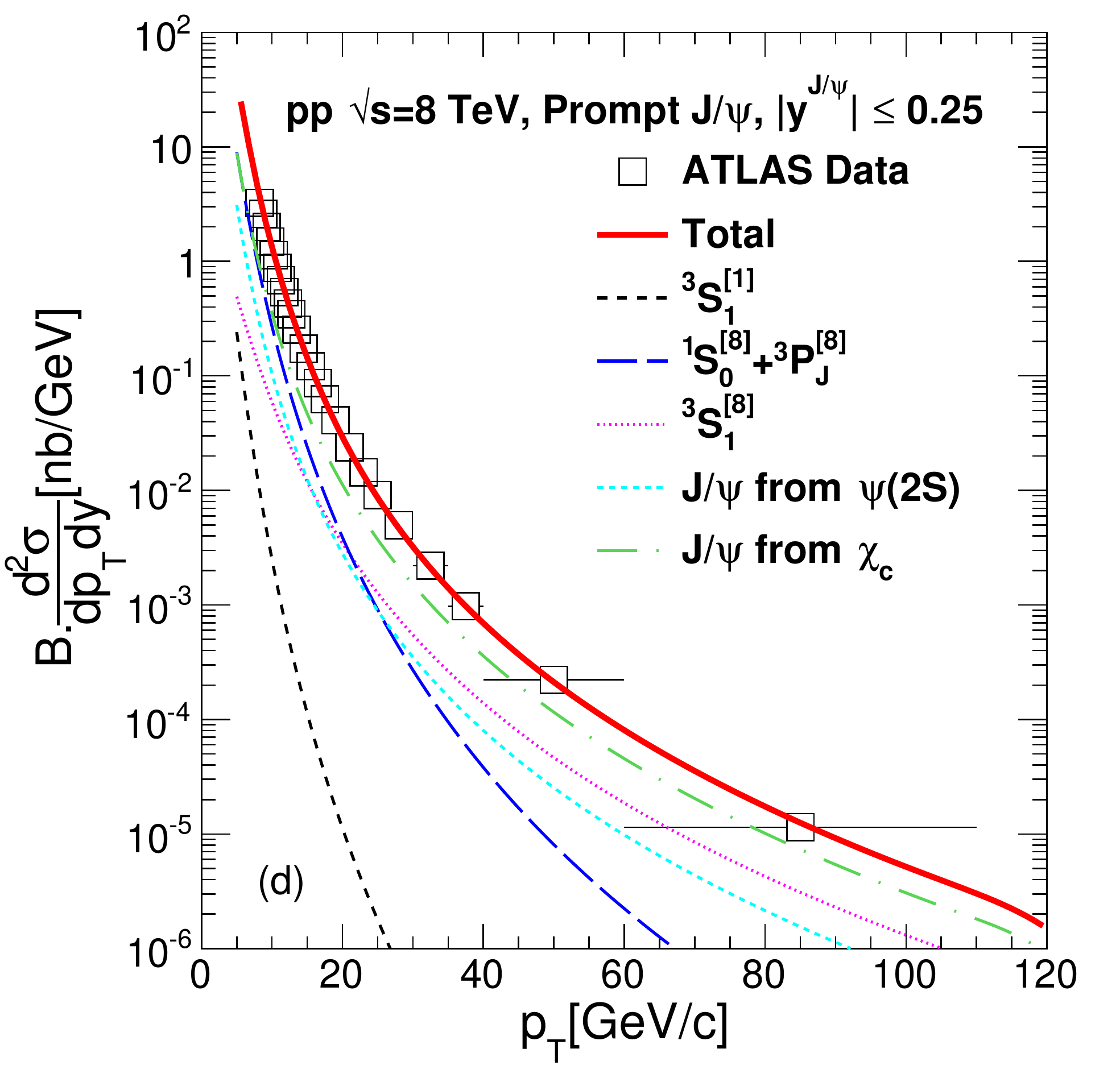}
\caption{(Color online) The NRQCD calculations of production cross section of $J/\psi$ in p+p collisions 
  as a function of transverse momentum compared with the measured data at LHC 
  (a) CMS data at $\sqrt{s}$ = 7 TeV~\cite{Chatrchyan:2011kc} 
  (b) CMS data at $\sqrt{s}$ = 7 TeV~\cite{Khachatryan:2015rra} 
  (c) ATLAS data at $\sqrt{s}$ = 7 TeV and
  (d) ATLAS data at $\sqrt{s}$ = 8 TeV~\cite{Aad:2015duc}. 
  The LDMEs are obtained by a combined fit of the LHC and Tevatron data.
}
\label{Fig:LDMEJPsi}
\end{figure}

\begin{figure}
\includegraphics[width=0.49\textwidth]{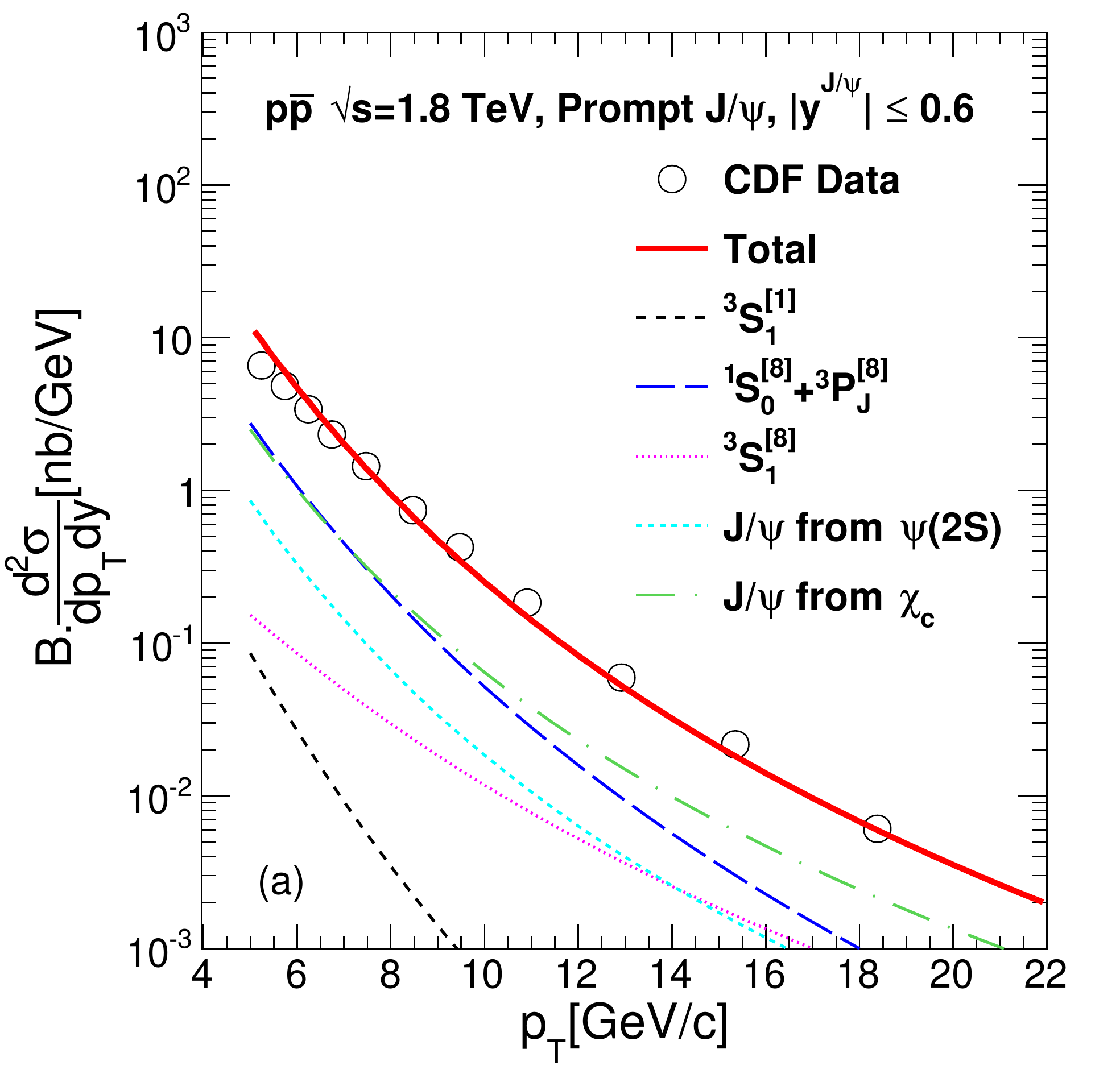}
\includegraphics[width=0.49\textwidth]{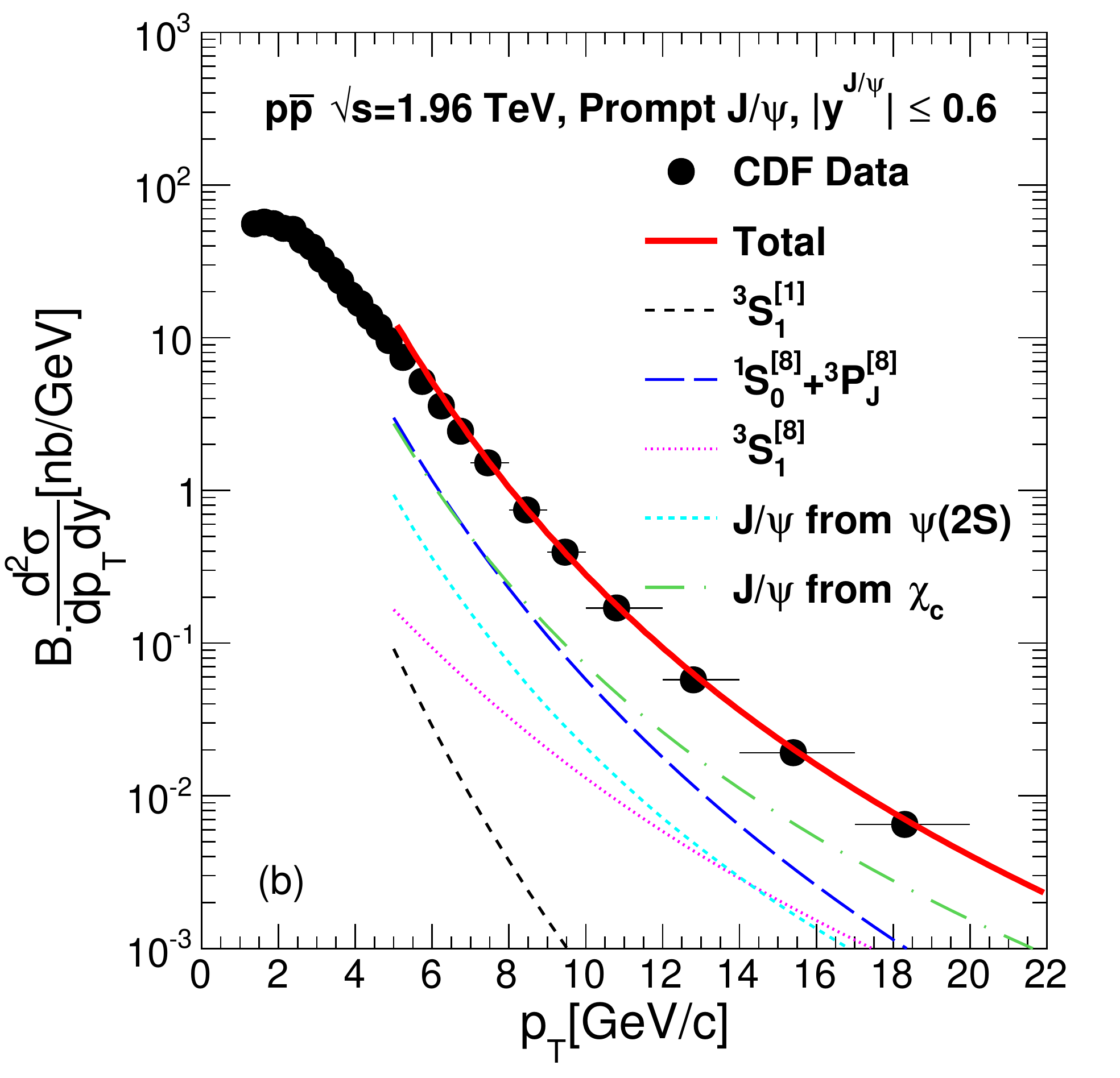}
\caption{(Color online) 
  The NRQCD calculations of production cross section of $J/\psi$ in p+${\bar {\rm p}}$ collisions 
  as a function of transverse momentum compared with the measured data at Tevatron 
  (a) CDF data at $\sqrt{s}$ = 1.8 TeV~\cite{Abe:1997jz} and
  (b) CDF data at $\sqrt{s}$ = 1.96 TeV~\cite{Acosta:2004yw}. 
  The LDMEs are obtained by a combined fit of the LHC and Tevatron data.
}
\label{Fig:LDMEJPsiCDF}
\end{figure}

\begin{figure}
\includegraphics[width=0.49\textwidth]{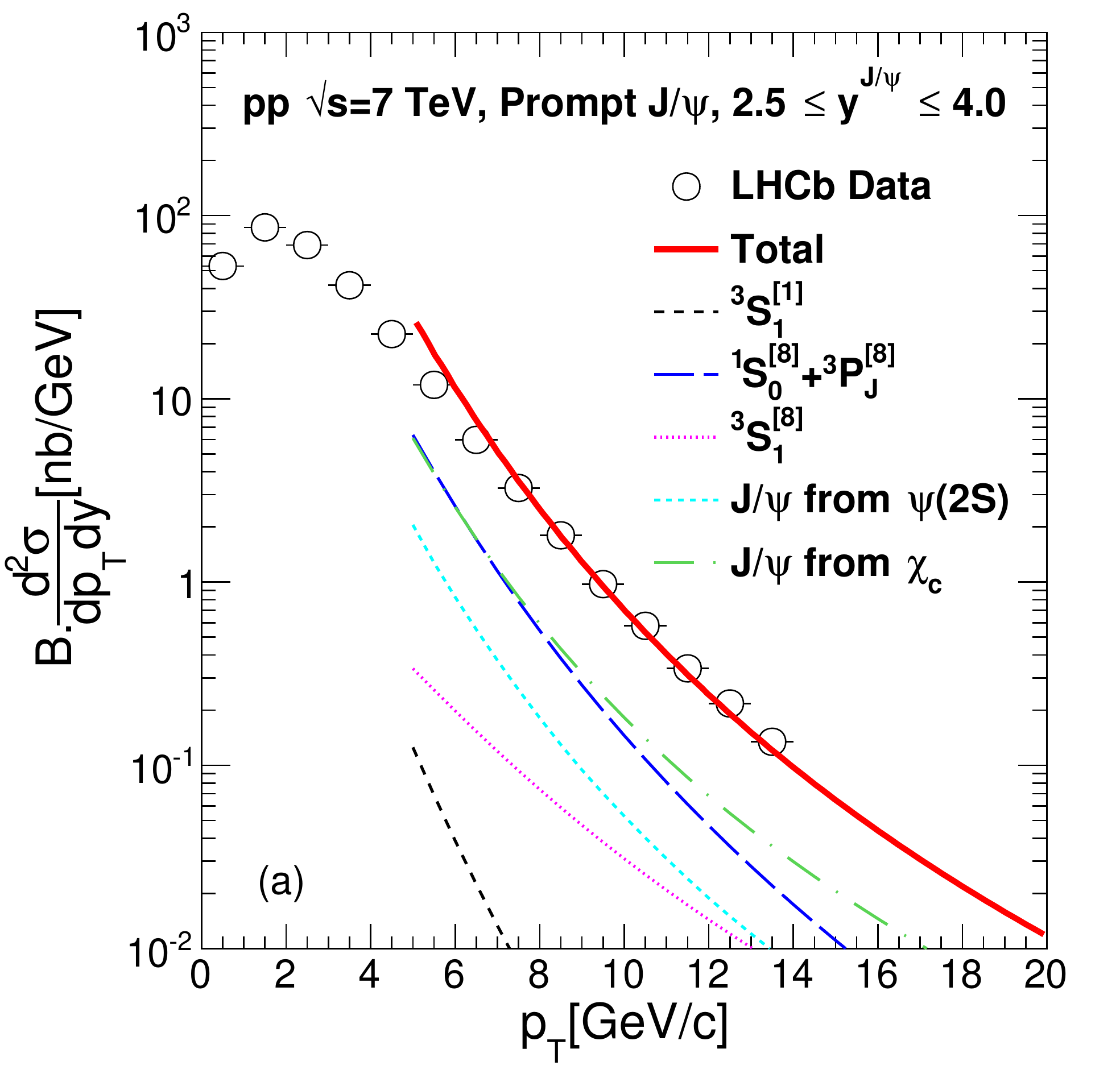}
\includegraphics[width=0.49\textwidth]{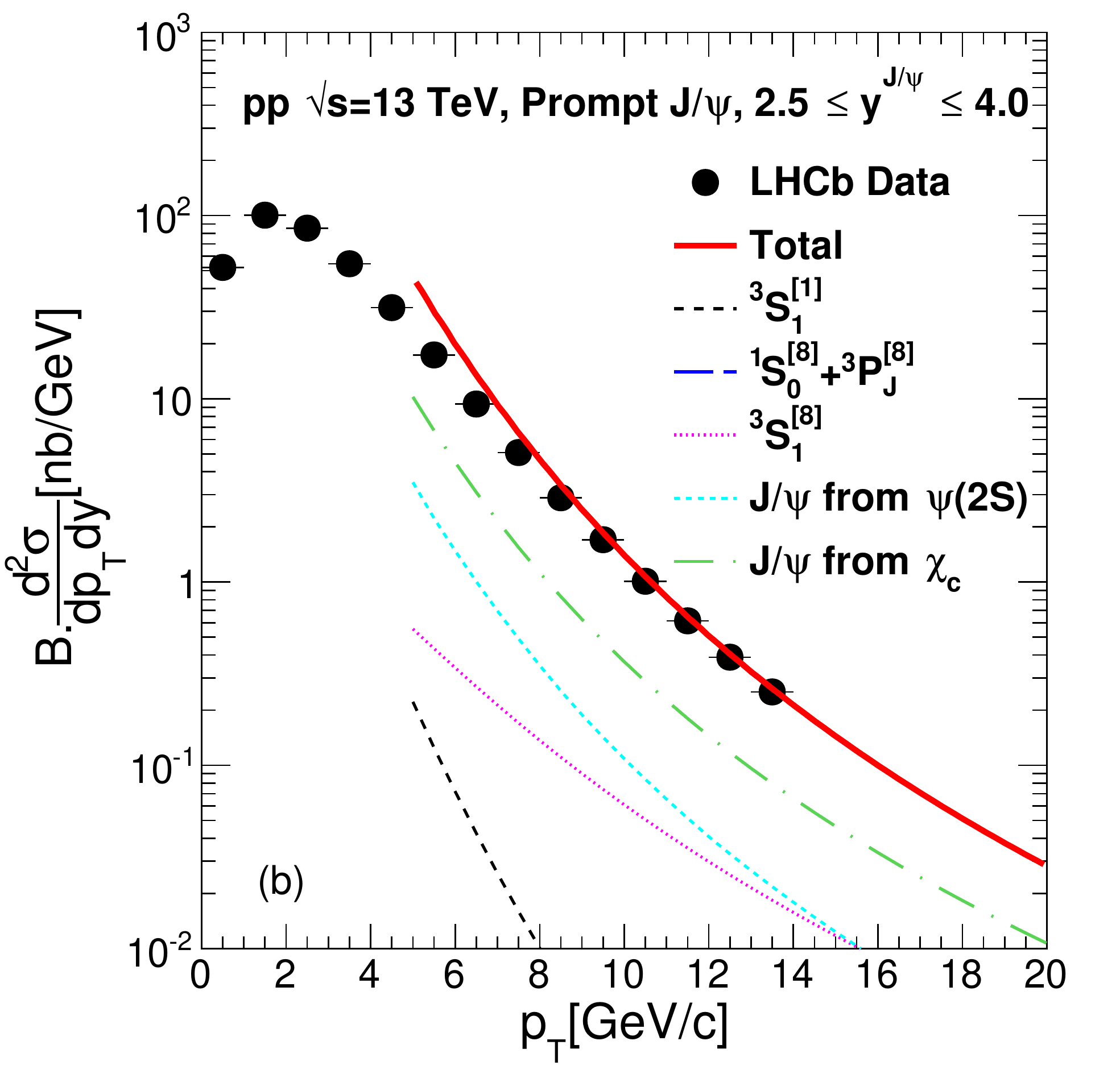}
\caption{(Color online) The NRQCD calculations of production cross section of $J/\psi$ in p+p collisions 
  as a function of transverse momentum compared with the measured data at LHC 
  (a) LHCb data at $\sqrt{s}$ = 7 TeV~\cite{Aaij:2011jh} and 
  (b) LHCb data at $\sqrt{s}$ = 13 TeV~\cite{Aaij:2015rla}. 
  The LDMEs are obtained by a combined fit of the LHC and Tevatron data.
}
\label{Fig:LDMEJPsiLHCb}
\end{figure}


\begin{figure}
\includegraphics[width=0.49\textwidth]{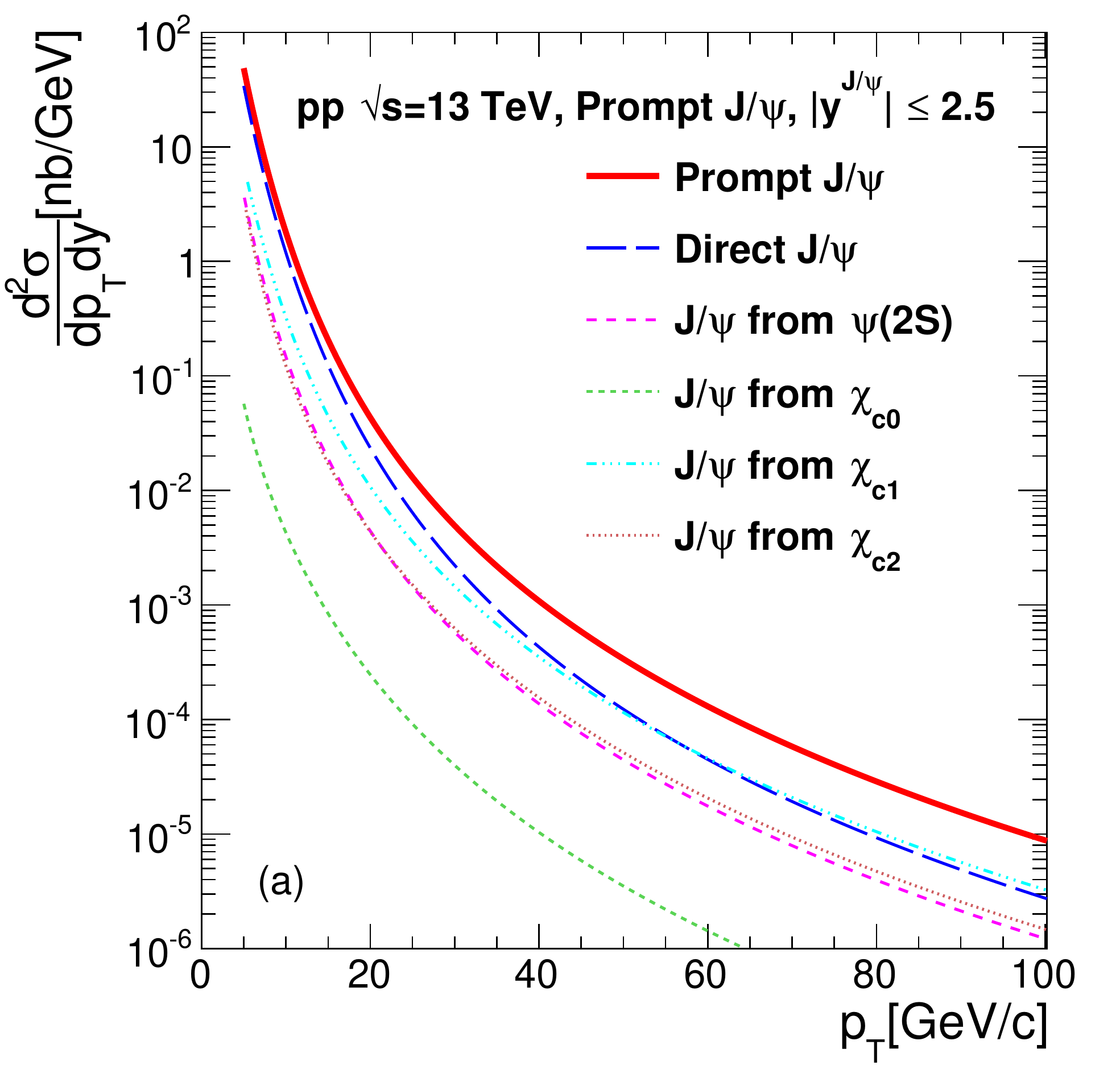}
\includegraphics[width=0.49\textwidth]{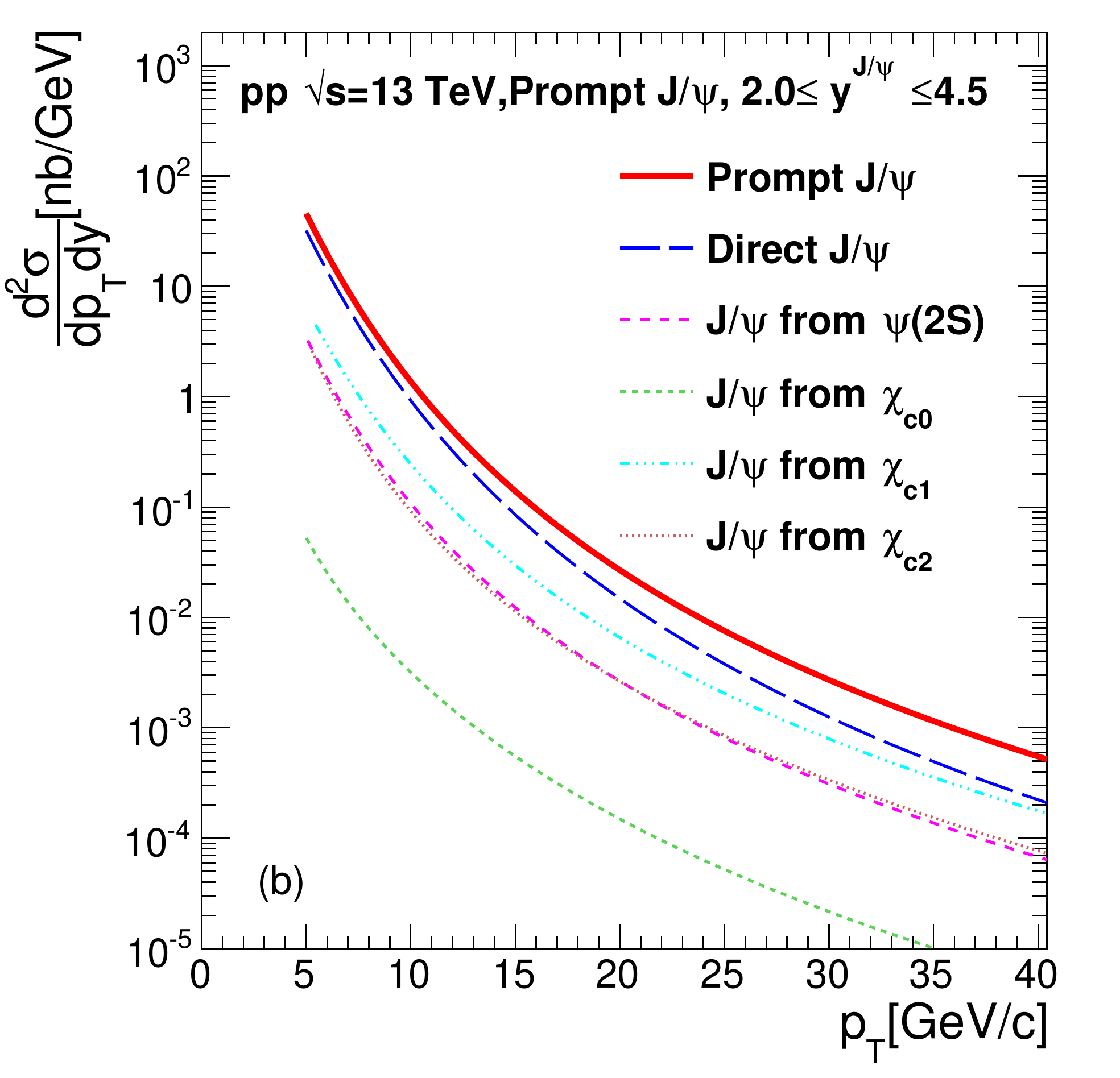}
\caption{(Color online) The NRQCD calculations of production cross section of $J/\psi$ in p+p collisions 
  as a function of transverse momentum at $\sqrt{s}$ = 13 TeV. The calculations are shown in the kinematic
  bins relevant to (a) CMS, ATLAS and (b) ALICE, LHCb detectors at LHC. For the J/$\psi$ meson all the 
  relevant contributions from higher mass states are also shown.}
\label{Fig:SigmaJPsi}
\end{figure}

\begin{figure}
\includegraphics[width=0.49\textwidth]{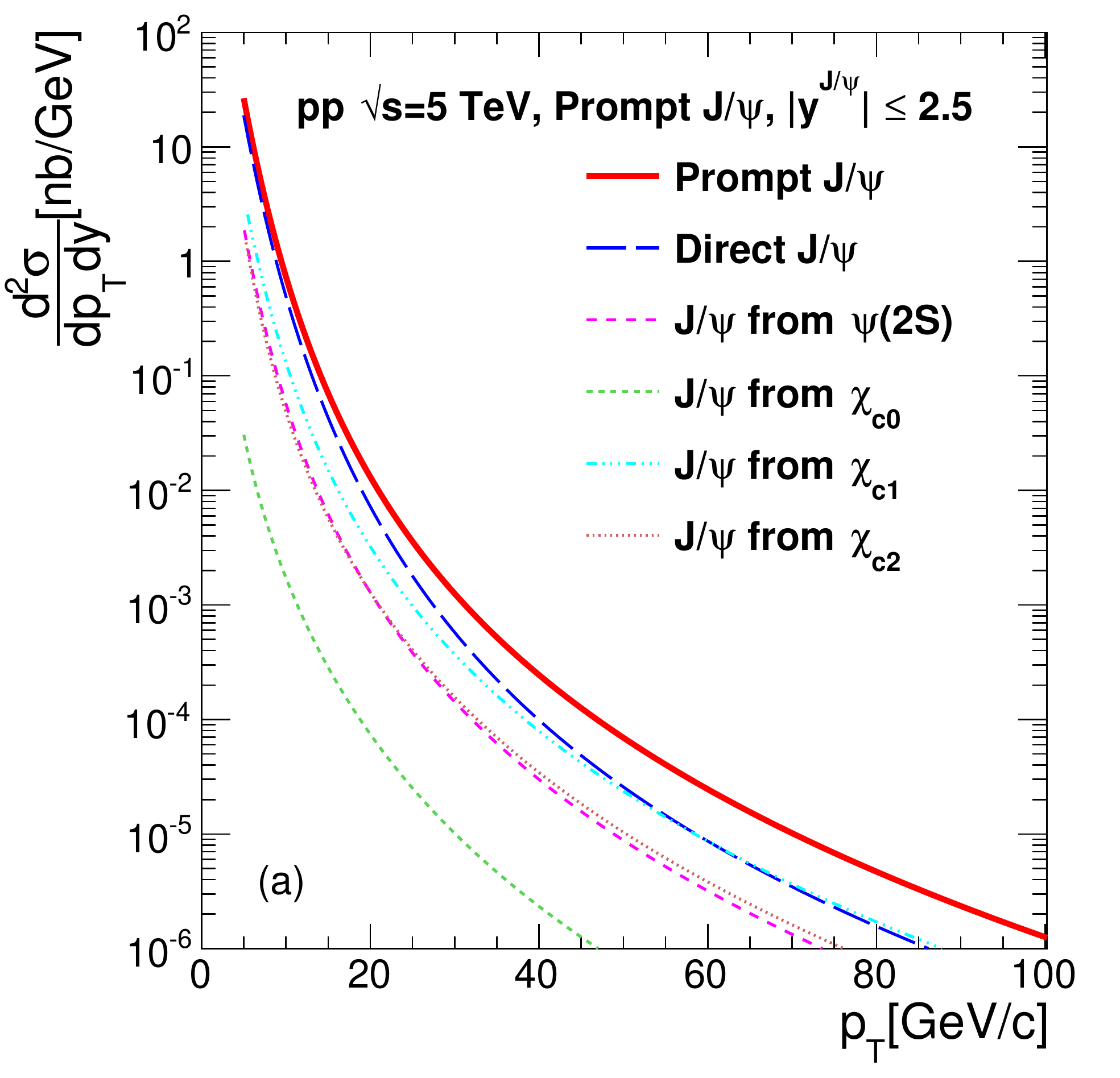}
\includegraphics[width=0.49\textwidth]{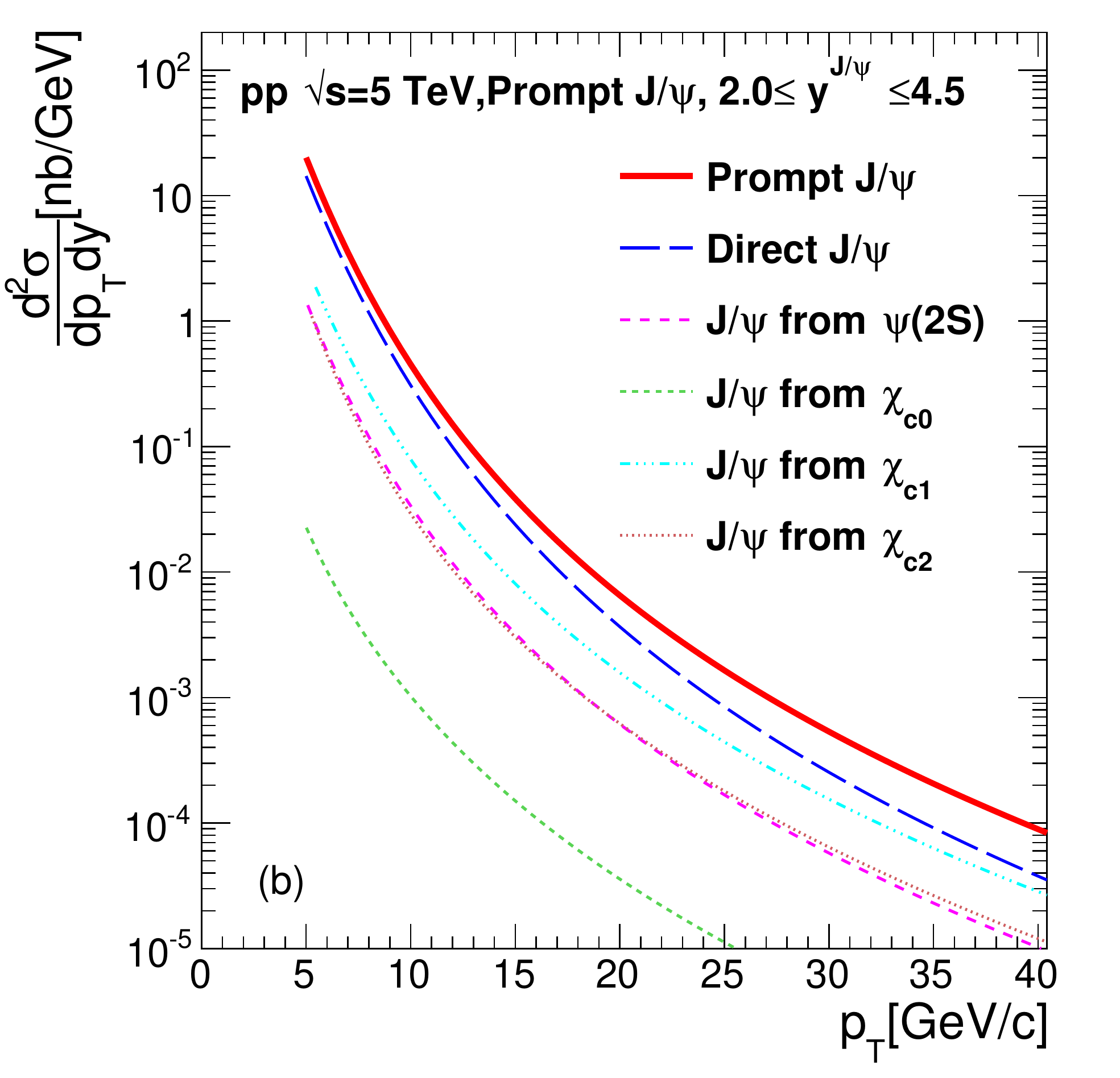}
\caption{(Color online)The NRQCD calculations of production cross section of $J/\psi$ in p+p collisions 
  as a function of transverse momentum at $\sqrt{s}$ = 5 TeV. The calculations are shown in the kinematic
  bins relevant to (a) CMS, ATLAS and (b) ALICE, LHCb detectors at LHC. For the J/$\psi$ meson all the 
  relevant contributions from higher mass states are also shown.
}
\label{Fig:SigmaJPsi5TeV}
\end{figure}


\section{Results and Discussions}
As discussed in the last section there are two free parameters ($M_{L}(c\barc([^3S_1]_{8}\rightarrow J/\psi))$,
 $M_{L}(c\barc([^1S_0]_{8},[^3P_0]_{8})\rightarrow J/\psi)$) for $J/\psi$, 
two ($M_{L}(c\barc([^3S_1]_{8}\rightarrow \psi(2S)))$, $M_{L}(c\barc([^1S_0]_{8},[^3P_0]_{8})\rightarrow \psi(2S))$)  
for $\psi$(2S) and one $(M_{L}(c\barc([^3S_1]_{8})\rightarrow \chi_{c0}))$ for $\chi_{cJ}$ to be obtained 
from the experiments. 
The measured yields of $\chi_{cJ}$ from the following datasets
are used to obtain color-octet matrix elements for $\chi_{cJ}$ 
\begin{enumerate}
\item{CDF results at $\sqrt{S}=1.8$~TeV~\cite{Abe:1997yz}}.
\item{ATLAS results at $\sqrt{S}=7$~\cite{ATLAS:2014ala}}.
\item{CMS results at $\sqrt{S}=7$~TeV~\cite{Chatrchyan:2012ub}}.
\item{LHCb results at $\sqrt{S}=7$~TeV~\cite{Aaij:2013dja}}.
\end{enumerate}
Figure~\ref{Fig:LDMEChicATLAS} shows the NRQCD calculations of production cross section 
of (a) $\chi_{c1}$, (b) $\chi_{c2}$ in p+p collisions at
$\sqrt{s}$ = 7 TeV and (c) $J/\psi$ from $\chi_{c1}$ and $\chi_{c2}$ 
decays in p+${\bar {\rm p}}$ collisions at $\sqrt{s}$ = 1.8 TeV as 
a function of transverse momentum. The calculations are compared with 
the measured data by ATLAS experiment at LHC~\cite{ATLAS:2014ala}
and measured data by CDF experiment at Tevatron~\cite{Abe:1997yz}. 
The $\chi_{c}$ color octet LDMEs are obtained by fitting this data.
Figure~\ref{Fig:LDMEChicCMS_LHCb} shows
the NRQCD calculations of production cross section ratios 
of $\chi_{c2}$ and $\chi_{c1}$ in p+p collisions at
$\sqrt{s}$ = 7 TeV as a function of transverse momentum. 
The calculations are compared with the measured data at
LHC in panel (a) CMS data at $\sqrt{s}$ = 7 TeV~\cite{Chatrchyan:2012ub}
and in panel (b) LHCb data at $\sqrt{s}$ = 7 TeV~\cite{Aaij:2013dja}.  
The $\chi_{c}$ color octet LDMEs are obtained by combined fitting of these datasets
and its value is 
\begin{equation}
  \begin{split}
    M_{L}(Q\barQ([^3S_1]_{8})\rightarrow \chi_{c0}) & 
    = (0.01112\pm {\color{black} 0.00068})\,{\rm GeV^3}\;,
    \label{eq:Mchicn1CO_val}
  \end{split}
\end{equation}
with a combined {\color{black} $\chi^2/dof=1.20$}. 

The measured yields of prompt $\psi(2S)$ from the following datasets
are used to obtain color-octet matrix elements for $\psi(2S)$ 
\begin{enumerate}
\item{CMS results at $\sqrt{S}=7$~TeV~\cite{Chatrchyan:2011kc,Khachatryan:2015rra}}.
\item{ATLAS results at $\sqrt{S}=7$ and 8 ~TeV~\cite{Aad:2015duc}}.
\item{CDF results at $\sqrt{S}=1.8$~TeV~\cite{Abe:1997jz}}.
\item{CDF results at $\sqrt{S}=1.96$~TeV~\cite{Acosta:2004yw}}.
\item{LHCb results at $\sqrt{S}=7$~TeV~\cite{Aaij:2012ag}}.
\end{enumerate}
Figure~\ref{Fig:LDMEPsi2S} shows the NRQCD calculations of production cross section of 
$\psi$(2S) in p+p collisions as a function of transverse momentum compared with the 
measured data at LHC in panels
(a) CMS data at $\sqrt{s}$ = 7 TeV~\cite{Chatrchyan:2011kc},
(b) CMS data at $\sqrt{s}$ = 7 TeV~\cite{Khachatryan:2015rra}, 
(c) ATLAS data at $\sqrt{s}$ = 7 TeV and, 
(d) ATLAS data at $\sqrt{s}$ = 8 TeV~\cite{Aad:2015duc}. 
 

 Figure~\ref{Fig:LDMEPsi2SCDF} shows the NRQCD calculations of production cross section 
of $\psi$(2S) in p+${\bar {\rm p}}$ and p+p collisions as a function of transverse 
momentum compared with the measured  data in panels
(a) CDF data at $\sqrt{s}$ = 1.8 TeV~\cite{Abe:1997jz}, 
(b) CDF data at $\sqrt{s}$ = 1.96 TeV~\cite{Acosta:2004yw} and 
(c) LHCb data at $\sqrt{s}$ = 7 TeV~\cite{Aaij:2012ag}.
 The LDMEs are obtained by a combined fit of the Tevatron and
LHC data

We obtain following values of $\psi(2S)$ color-octet matrix elements by a combined fit of 
the Tevatron and LHC data   

\begin{equation}
  \begin{split}
    M_{L}(c\barc([^3S_1]_{8})\rightarrow \psi(2S)) &= (0.00362\pm 0.00006 \pm {\color{black} 0.00002}) \, {\rm GeV^3}\\
    M_{L}(Q\barQ([^1S_0]_{8},[^3P_0]_{8})\rightarrow \psi(2S)) &= (0.02280\pm 0.00028 \pm {\color{black}0.00034}) \,{\rm GeV^3}\\
     ~\label{eq:COLDME_Psi}
  \end{split}
\end{equation}
with a {\color{black}$\chi^2/dof=2.54$}.

{\color{black}
 Here the first error is due to fitting and the second error is 
 obtained by enhancing the CS cross section 3 times. 
 It is due to the fact that NLO corrections enhance the total color-singlet J/$\psi$ 
 production by a factor of 2~\cite{Gong:2008sn}.
 The NLO corrections to J/$\psi$ production via S-wave color octet (CO) states 
 ($^1S_{0}^{[8]}\,^3S_{1}^{[8]}$) are found to be small Ref.~\cite{Gong:2008ft}.
}
To fit the remaining 2 parameters of $J/\psi$ we use the combined fit for the
following datasets of prompt $J/\psi$ yields
\begin{enumerate}
\item{CMS results at $\sqrt{S}=7$~TeV~\cite{Chatrchyan:2011kc,Khachatryan:2015rra}}.
\item{ATLAS results at $\sqrt{S}=7$ and 8 ~TeV~\cite{Aad:2015duc}}.
\item{CDF results at $\sqrt{S}=1.8$~TeV~\cite{Abe:1997jz}}.
\item{CDF results at $\sqrt{S}=1.96$~TeV~\cite{Acosta:2004yw}}.
\item{LHCb results at $\sqrt{S}=7$~TeV~\cite{Aaij:2011jh}}.
\item{LHCb results at $\sqrt{S}=13$~TeV~\cite{Aaij:2015rla}}.
\end{enumerate}

 Figures~\ref{Fig:LDMEJPsi} shows the NRQCD calculations of production cross section of 
$J/\psi$ in p+p collisions as a function of transverse momentum compared with 
the measured data at LHC in panels (a) CMS data at $\sqrt{s}$ = 7 TeV~\cite{Chatrchyan:2011kc} 
and (b) CMS data at $\sqrt{s}$ = 7 TeV~\cite{Khachatryan:2015rra} (c) ATLAS data at $\sqrt{s}$ = 7 TeV 
and (d) ATLAS data at $\sqrt{s}$ = 8 TeV~\cite{Aad:2015duc}. 
   Figure~\ref{Fig:LDMEJPsiCDF} shows  the NRQCD calculations of production cross 
section of $J/\psi$ in p+{$\bar {\rm p}$}  collisions as compared with the measured data at 
Tevatron in panels 
(a) CDF data at $\sqrt{s}$ = 1.8 TeV~\cite{Abe:1997jz} and 
(b) CDF data at $\sqrt{s}$ = 1.96 TeV~\cite{Acosta:2004yw}. 
  Figure~\ref{Fig:LDMEJPsiLHCb} shows the the NRQCD calculations of production cross 
section of $J/\psi$ in p+p collisions compared with the forward rapidity data measured at LHC in panels
(a) LHCb data at $\sqrt{s}$ = 7 TeV~\cite{Aaij:2011jh} and (b) LHCb data at $\sqrt{s}$ = 13 TeV 
~\cite{Aaij:2015rla}. 
 We obtain following values of $J/\psi$ color-octet matrix elements by a combined fit of 
the Tevatron and the LHC data

\begin{equation}
\begin{split}
M_{L}(c\barc([^3S_1]_{8})\rightarrow J/\psi)&= (0.00206\pm 0.00014 \pm {\color{black}0.00001}) \,{\rm GeV^3}\\
 M_{L}(Q\barQ([^1S_0]_{8},[^3P_0]_{8})\rightarrow J/\psi)&= (0.06384\pm 0.00106 \pm {\color{black}0.00062}) \,{\rm GeV^3} \\
~\label{eq:COLDME_JPsi}
\end{split}
\end{equation}
with a {\color{black}$\chi^2/dof=2.76$}.

\begin{table}[h]
\caption{Comparison of $\chi_{c0}$ LDMEs. The short distance calculations are at LO except Ref.~\cite{Jia:2014jfa}(NLO).}
\begin{tabular}{|l|c|c|c|c|}
\hline            
Ref.                             &PDF     &m$_{c}$      &$M_{L}(c\barc([^3P_0]_{1})\rightarrow \chi_{c0})$                     &$M_{L}(c\barc([^3S_1]_{8})\rightarrow \chi_{c0})$      \\        
                                &        &(GeV)       &(GeV$^5$)                        &(GeV$^3$)         \\
\hline
ours                           &CTEQ6M   &1.6         &0.054m$_{c}^{2}$                 &0.01112$\pm$0.00068\\
\cite{Cho:1995vh}              &MRSD0    &1.48        &$--$                            &0.0098$\pm$0.0013   \\
\cite{Braaten:1999qk}          &MRST98LO &1.5         &0.089$\pm$0.013                 &0.0023$\pm$0.0003    \\
\cite{Braaten:1999qk}          &CTEQ5L   &1.5         &0.091$\pm$0.013                 &0.0019$\pm$0.0002     \\
\cite{Sharma:2012dy}           &MSTW08LO &1.4         &0.054m$_{c}^{2}$                   &0.00187$\pm$0.00025     \\
\cite{Jia:2014jfa}(LO)         &CTEQ6L   &1.5         &$--$                            &0.00031$\pm$0.00009   \\ 
\cite{Jia:2014jfa}(NLO)        &CTEQ6M   &1.5         &$--$                            &0.0021$\pm$0.00004  \\ 
\hline
\end{tabular}
\label{table:LDMEChic0}
\end{table}

\begin{table}[h]
\caption{Comparison of $\psi$(2S) LDMEs. The short distance calculations are at LO. }
\begin{tabular}{|l|c|c|c|c|c|c|}
\hline            
Ref.       &PDF     &m$_{c}$  &$M_{L}(c\barc([^3S_1]_{1}$      &$M_{L}(c\barc([^3S_1]_{8}$    &$M_{L}(c\barc([^1S_0]_{8},$ \\
           &        &        &$\rightarrow \psi(2S)))$      &$\rightarrow \psi(2S)))$    &$[^3P_0]_{8})\rightarrow \psi(2S)))$   \\
           &        &(GeV)   &(GeV$^{3}$)                    &(GeV$^3$)                   &(GeV$^3$)                            \\ 
\hline
ours                            &CTEQ6M   &1.6     &0.76           &0.00362$\pm$0.00006 &0.02280$\pm$0.00028                                \\
\cite{Cho:1995vh}               &MRSD0    &1.48    &$--$           &0.0046$\pm$0.0010   &0.0059$\pm$0.0019                                 \\
\cite{Braaten:1999qk}           &MRST98LO &1.5     &0.65$\pm$0.6   &0.0042$\pm$0.0010   &0.0037$\pm$0.0014                                 \\
\cite{Braaten:1999qk}           &CTEQ5L   &1.5     &0.67$\pm$0.7   &0.0037$\pm$0.0090   &0.0022$\pm$0.001                                  \\
\cite{Sharma:2012dy}            &MSTW08LO &1.4     &0.76           &0.0033$\pm$0.00021  &0.01067$\pm$0.0009                               \\
\cite{Beneke:1996yw}            &CTEQ4L   &1.5     &$--$           &0.0044$\pm$0.0008   &0.00514$\pm$0.0016                               \\
\cite{Beneke:1996yw}            &GRV94LO  &1.5     &$--$           &0.0046$\pm$0.0008   &0.00457$\pm$0.0014                                \\
\cite{Beneke:1996yw}            &MRSR2    &1.5     &$--$           &0.0056$\pm$0.0011   &0.01246$\pm$0.0027                                \\
\hline
\end{tabular}
\label{table:LDMEPsi2S}
\end{table}

\begin{table}[h]
\caption{Comparison of J/$\psi$ LDMEs. The short distance calculations are at LO except Ref.~\cite{Butenschoen:2010rq}.}
\begin{tabular}{|l|c|c|c|c|c|c|}
\hline            
Ref.       &PDF     &m$_{c}$  &$M_{L}(c\barc([^3S_1]_{1}$      &$M_{L}(c\barc([^3S_1]_{8}$    &$M_{L}(c\barc([^1S_0]_{8}$ \\
           &        &        &$\rightarrow J/\psi))$      &$\rightarrow J/\psi))$    &$,[^3P_0]_{8})\rightarrow J/\psi))$   \\
          &        &(GeV)   &(GeV$^{3}$)            &(GeV$^{3}$)   &(GeV$^{3}$)                                  \\
\hline
ours                           &CTEQ6M   &1.6     &1.2            &0.00206$\pm$0.00014  &0.06384$\pm$0.00106                           \\
\cite{Cho:1995vh}              &MRSD0    &1.48    &$--$           &0.0066$\pm$0.0021    &0.0220$\pm$0.050                                 \\
\cite{Braaten:1999qk}          &MRST98LO &1.5     &1.3$\pm$0.1    &0.0044$\pm$0.0007    &0.026$\pm$0.0026                                \\
\cite{Braaten:1999qk}          &CTEQ5L   &1.5     &1.4$\pm$0.1    &0.0039$\pm$0.0007    &0.0194$\pm$0.0021                                \\
\cite{Sharma:2012dy}           &MSTW08LO &1.4      &1.2            &0.0013$\pm$0.0013    &0.0239$\pm$0.0115                                \\
\cite{Beneke:1996yw}            &CTEQ4L   &1.5     &$--$           &0.0106$\pm$0.0014   &0.0125$\pm$0.0032                                \\
\cite{Beneke:1996yw}            &GRV94LO  &1.5     &$--$           &0.0112$\pm$0.0014   &0.0114$\pm$0.0032                                \\
\cite{Beneke:1996yw}            &MRSR2    &1.5     &$--$           &0.0140$\pm$0.0022   &0.0311$\pm$0.0059                                \\
\cite{Butenschoen:2010rq}      &CTEQ6M   &1.5     &1.32           &0.00312$\pm$0.00093 &0.00962$\pm$0.0008                              \\
\hline
\end{tabular}
\label{table:LDMEJPsi}
\end{table}


Table~\ref{table:LDMEChic0} shows $\chi_{c0}$ LDMEs extracted in present analysis along with the results from other
analysis. The value of charm quark mass as well as the PDFs used in the calculations are also shown in the table.
The short distance calculations are at LO except the last row in the table. We have made major extension in fitting
the $\chi_{c}$ LDME. All the earlier calculations~\cite{Cho:1995vh,Braaten:1999qk,Sharma:2012dy} use only CDF data to
fit the $\chi_{c}$ LDME. We use CDF data~\cite{Abe:1997yz} along-with the data from LHC~\cite{ATLAS:2014ala,Chatrchyan:2012ub,Aaij:2013dja} 
to constrain the CO LDME of $\chi_{c}$. The new high energy LHC data require larger value of $(M_{L}(c\barc([^3S_1]_{8})\rightarrow \chi_{c0}))$ to fit the data.  

Table~\ref{table:LDMEPsi2S} shows $\psi$(2S) LDMEs extracted in present analysis along with the results from other 
works. All the calculations are at LO in $\alpha_{s}$. The calculations in Ref.~\cite{Cho:1995vh,Braaten:1999qk,Beneke:1996yw} use only 
CDF data to fit the LDMEs while the Ref.~\cite{Sharma:2012dy} uses CDF and LHC data at mid-rapidity. In our analysis we use data from
CDF~\cite{Abe:1997jz,Acosta:2004yw} and LHC data in mid rapidity~\cite{Chatrchyan:2011kc,Khachatryan:2015rra,Aad:2015duc} as well 
as LHC data in forward rapidity~\cite{Aaij:2012ag}, both the datasets covering a much wider $p_T$ range. Our value of matrix 
element $M_{L}(c\barc([^3S_1]_{8}\rightarrow \psi(2S)))$ is similar with other analysis. The value of linear-combination, 
$M_{L}(c\barc([^1S_0]_{8},[^3P_0]_{8})\rightarrow \psi(2S))$, varies significantly from 0.0022 to
0.01246 between different analysis. As it can be seen from Table~\ref{table:LDMEPsi2S}, the high energy LHC data require larger value of
$M_{L}(c\barc([^1S_0]_{8},[^3P_0]_{8})\rightarrow \psi(2S))$.    

Table~\ref{table:LDMEJPsi} shows J/$\psi$ LDMEs extracted in present analysis along with the results from other 
works. All the calculations except Ref.~\cite{Butenschoen:2010rq} are at LO in $\alpha_{s}$. 
The calculations in Ref.~\cite{Cho:1995vh,Braaten:1999qk,Beneke:1996yw} 
use only CDF data to fit the LDMEs while Ref.~\cite{Sharma:2012dy} uses CDF, RHIC and LHC data at mid-rapidity. 
In our analysis, we use data from CDF~\cite{Abe:1997jz,Acosta:2004yw} and LHC data in mid rapidity~\cite{Chatrchyan:2011kc,Khachatryan:2015rra,Aad:2015duc} 
as well as LHC data in forward rapidity~\cite{Aaij:2011jh,Aaij:2015rla}, both the datasets covering a much wider $p_T$ range.
The value of the matrix element $M_{L}(c\barc([^3S_1]_{8}\rightarrow J/\psi))$ is different in different analysis.
The large error present on $M_{L}(c\barc([^3S_1]_{8}\rightarrow J/\psi))$ in Ref.~\cite{Sharma:2012dy} is significantly improved by our 
simultaneous fitting of several datasets. The value of linear-combination, 
$M_{L}(c\barc([^1S_0]_{8},[^3P_0]_{8})\rightarrow J/\psi)$, varies significantly from 0.0114 to 
0.06384 (our value) between different analysis at LO. The NLO analysis~\cite{Butenschoen:2010rq} does not fit the linear combination but
fit both $M_{L}(c\barc([^1S_0]_{8}\rightarrow J/\psi))$ and $M_{L}(c\barc([^3P_0]_{8}\rightarrow J/\psi))$) 
LDMEs independently and their values are given as 0.0450$\pm$0.0072 and -0.0121$\pm$0.0035 respectively . The value of $M_{L}(c\barc([^1S_0]_{8},[^3P_0]_{8})\rightarrow J/\psi)$
is very small for Ref.~\cite{Butenschoen:2010rq} because of the negative value of $M_{L}(c\barc([^3P_0]_{8}\rightarrow J/\psi))$.


 We use our newly constrained CO LDMEs shown in equation~\ref{eq:COLDME_JPsi} to 
predict the J/$\psi$ cross-section at 13 TeV and 5 TeV for the kinematical 
bins relevant to LHC detectors.     
  Figure~\ref{Fig:SigmaJPsi} shows the NRQCD 
calculations of production cross section of $J/\psi$ in p+p collisions  
as a function of transverse momentum at $\sqrt{s}$ = 13 TeV. 
Figure~\ref{Fig:SigmaJPsi5TeV} is same as Fig.~\ref{Fig:SigmaJPsi} but at 
$\sqrt{s}$ = 5 TeV. 
 Both the figures give calculations in the kinematic bins relevant for (a) CMS, ATLAS and 
(b) ALICE, LHCb detectors at LHC. For the J/$\psi$ meson all the relevant contributions 
from higher mass states are also shown. 


\section{Summary}

  We have presented NRQCD calculations for the differential production 
  cross sections of prompt J/$\psi$ and prompt $\psi$(2S) in  p+p collisions.
  For the J/$\psi$ meson, all the relevant contributions from higher mass states 
  are estimated. 
  Measured transverse momentum distributions of $\psi$(2S), $\chi_{\rm c}$ and J/$\psi$ 
  in p +{$\bar {\rm p}$} collisions at $\sqrt{s}=$ 1.8, 1.96 TeV and in p+p collisions at 
  7, 8 and 13 TeV are used to constrain LDMEs. 
  The calculations for  prompt J/$\psi$ and prompt $\psi$(2S) are compared with the measured 
  data at Tevatron and LHC. 
  The formalism provides  very good description of the data in wide energy range. 
  The values of LDMEs are used to predict the charmonia cross sections in p+p collisions 
  at 13 and 5 TeV in kinematic bins relevant for LHC detectors. 
  We compare the LDMEs for charmonia obtained in this analysis with the results from earlier works.
 {\color{black}   
   At high $p_T$, the color singlet contribution is very small and thus the LHC data in large $p_T$ range 
   help to constrain the relative contributions of different colour octet contributions.
   The high energy LHC data require a smaller value of the LDME $M_{L}(c\barc([^3S_1]_{8}\rightarrow \psi))$ 
   and a larger value for the combination  $M_{L}(c\barc([^1S_0]_{8},[^3P_0]_{8})\rightarrow \psi)$ of LDMEs.   
   In summary, we present a comprehensive lowest-order analysis of hadroproduction data, 
   including very recent LHC data. The values of fitted LDMEs will be useful for 
   predictions of quarkonia cross section and for the purpose of a comparison with those 
   obtained using NLO formulations.
 }

\section*{Acknowledgement}
 We acknowledge the fruitful discussions on this topic with Rishi Sharma.

\section*{References}

\end{document}